\documentclass[journal]{IEEEtran}
\usepackage{amsmath}
\usepackage{graphicx}
\usepackage{caption2}
\usepackage{amsthm}
\usepackage{float}
\usepackage{verbatim}
\usepackage{epstopdf}
\usepackage{amssymb}
\usepackage{amsfonts}
\usepackage{subfigure}
\begin{document}
\title{On the Capacity of Vector Gaussian Channels With Bounded Inputs}
\author{{Borzoo Rassouli and Bruno Clerckx} %
\thanks{Borzoo Rassouli is with the Intelligent Systems and Networks group of Department of Electrical and Electronics,
Imperial College London, United Kingdom. email: b.rassouli12@imperial.ac.uk}
\thanks{Bruno Clerckx is with the Communication and Signal Processing group of Department of Electrical and Electronics,
Imperial College London and the School of Electrical Engineering, Korea University, Korea. email: b.clerckx@imperial.ac.uk}
\thanks{This paper was presented in part at the IEEE International Conference on Communications (ICC) 2015, London, UK.}
\thanks{This work was partially supported by the Seventh Framework Programme for Research of the European Commission under grant number HARP-318489.}}
\maketitle
\begin{abstract}
The capacity of a deterministic multiple-input multiple-output (MIMO) channel under the peak and average power constraints is investigated. For the identity channel matrix, the approach of Shamai et al. is generalized to the higher dimension settings to derive the necessary and sufficient conditions for the optimal input probability density function. This approach prevents the usage of the identity theorem of the holomorphic functions of several complex variables which seems to fail in the multi-dimensional scenarios. It is proved that the support of the capacity-achieving distribution is a finite set of hyper-spheres with mutual independent phases and amplitude in the spherical domain. Subsequently, it is shown that when the average power constraint is relaxed, if the number of antennas is large enough, the capacity has a closed form solution and constant amplitude signaling at the peak power achieves it. Moreover, it will be observed that in a discrete-time memoryless Gaussian channel, the average power constrained capacity, which results from a Gaussian input distribution, can be closely obtained by an input where the support of its magnitude is a discrete finite set. Finally, we investigate some upper and lower bounds for the capacity of the non-identity channel matrix and evaluate their performance as a function of the condition number of the channel.
\end{abstract}
\section{Introduction}
The capacity of a point-to-point communication system subject to peak and average power constraints was investigated in \cite{Smith} for the scalar Gaussian channel where it was shown that the capacity-achieving distribution is unique and has a probability mass function with a finite number of mass points. In \cite{Shamai}, Shamai and Bar-David gave a full account on the capacity of a quadrature Gaussian channel under the aforementioned constraints and proved that the optimal input distribution has a discrete amplitude and a uniform independent phase. This discreteness in the optimal input distribution was surprisingly shown in \cite{Abou} to be true even without a peak power constraint for the Rayleigh-fading channel when no channel state information (CSI) is assumed either at the receiver or the transmitter. Following this work, the authors in \cite{Katz} and \cite{Gursoy} investigated the capacity of noncoherent AWGN and Rician-fading channels, respectively. In \cite{Tchamkerten}, a point to point real scalar channel is considered in which sufficient conditions for the additive noise are provided such that the support of the optimal bounded input has a finite number of mass points. These sufficient conditions are also useful in multi-user settings as shown in \cite{khandani} for the MAC channel under bounded inputs.

The analysis of the MIMO channel under the peak power constraints per antenna is a straightforward problem after changing the vector channel into parallel AWGN channels and applying the results of \cite{Smith} or \cite{Shamai}. Recently, the vector Gaussian channel under the peak and average power constraints has become more practical by the new scheme proposed in \cite{Sedaghat}. More specifically, this scheme enables multiple antenna transmission using only one RF chain and the peak power constraint (i.e., a peak constraint on the norm of the input vector rather than on each antenna separately) is the very result of this single RF chain. The capacity of the vector Gaussian channel under the peak and average power constraints has been explored in \cite{Palanki} and \cite{Chan}. However, according to \cite{Sommerfeld}, it seems that the results in the higher dimension settings are not rigorous due to the usage of the identity theorem for holomorphic functions of several complex variables without fulfilling its conditions. As shown by an example in section IV of \cite{Sommerfeld}, a holomorphic function of several complex variables can be zero on $\mathbb{R}^n$, but not necessarily zero on $\mathbb{C}^n$. Since $\mathbb{R}^n$ is not an open subset of $\mathbb{C}^n$, the identity theorem cannot be applied. To address this problem, the contributions of this paper are as follows.

\begin{itemize}
  \item For the identity channel matrix, the approach of \cite{Shamai} is generalized to the vector Gaussian channel in which the complex extension will be done only on a single variable which is the amplitude of the input in the spherical coordinates. The necessary and sufficient conditions for the optimality of the input distribution are derived and it is proved that the magnitude of the capacity-achieving distribution has a probability mass function over a finite number of mass points which determines a finite number of hyper spheres in the spherical coordinates. Further, the magnitude and the phases of the capacity-achieving distribution are mutually independent and the phases are distributed in a way that the points are uniformly distributed on each of the hyper spheres.
  \item It is shown that if the average power constraint is relaxed, when the ratio of peak power to the number of dimensions remains below a certain threshold ($\approx 3.4$), the constant amplitude signaling at the peak power achieves the capacity.
  \item It is also shown that for a fixed SNR, the gap between the Shannon capacity and the constant amplitude signaling decreases as $O(\frac{1}{n})$ for large values of $n$, where $n$ denotes the number of dimensions.
  \item Finally, the case of the non-identity channel matrix is considered where we start from the MISO channel and show that the support of the optimal input does not necessarily have discrete amplitude. Afterwards, several upper bounds and lower bounds are provided for the general $n$ by $m$ MIMO channel capacity. The performance of these bounds are evaluated numerically as a function of the condition number of the channel.
\end{itemize}


The paper is organized as follows. The system model and some preliminaries are provided in section \ref{s2.1}, respectively. The main result of the paper is given in section \ref{s2.2} for the identity channel. The general case of the non-identity channel matrix is briefly investigated in section \ref{s3}. Numerical results and the conclusion are given in sections \ref{s4} and \ref{s5}, respectively. Some of the calculations are provided in the appendices at the end of the paper.
\section{System Model and preliminaries}\label{s2.1}
In a discrete-time memoryless vector Gaussian channel, the input-output relationship for the identity channel is given by
\begin{equation}\label{e1}
    \mathbf{Y}(t)=\mathbf{X}(t)+\mathbf{W}(t),
\end{equation}
where $\mathbf X(t)$, $\mathbf Y(t)$ ($\in \mathbb{R}^n$) denote the input and output of the channel, respectively. $t$ denotes the channel use and $\{\mathbf{W(t)}\}$ is an i.i.d. noise vector process with $\mathbf{W}(t)\sim N(\mathbf 0,\mathbf{I}_n)$ which is independent of $\mathbf{X}(t)$ for every transmission $t$. \footnote{It is obvious that the $m$-dimensional complex AWGN channel can be mapped to the channel in (\ref{e1}) with $n=2m.$}

The capacity of the channel in (\ref{e1}) under the peak and the average power constraints is
\begin{equation}\label{e2}
    C(u_p,u_a)=\sup_{F_{\mathbf X}(\mathbf x):\|\mathbf X\|^2\leq u_p,\  E(\|\mathbf X\|^2)\leq u_a}I(\mathbf X;\mathbf Y),
\end{equation}
where $F_{\mathbf X}(\mathbf x)$ denotes the input cumulative distribution function (CDF) of the input vector, and $u_p$, $u_a$ are the upper bounds for the peak and the average power, respectively. Throughout the paper, any operator that involves a random variable reads with the term \textit{almost-surely} (e.g. $\|\mathbf X\|^2\stackrel{a.s.}\leq u_p$)\footnote{More precisely, let $\Omega$ be the sample space of the probability model over which the random vector $\mathbf X$ is defined. $\|\mathbf{X}\|^2\stackrel{a.s.}\leq u_p$ is equivalent to $\mbox{Pr}\{\omega\in\Omega|\ \|\mathbf{X}(\omega)\|^2\leq u_p\}=1.$}.

It is obvious that
\begin{equation*}
    \sup_{F_{\mathbf X}(\mathbf x):\|\mathbf X\|^2\leq u_p,\  E(\|\mathbf X\|^2)\leq u_a}\!\!\!\!\!\!\!\!\!\!\!\!\!\!I(\mathbf X;\mathbf Y)\leq\!\!\! \sup_{F_{\mathbf X}(\mathbf x):E(\|\mathbf X\|^2)\leq \min (u_p,u_a)}\!\!\!\!\!\!\!\!\!\!\!\!\!\!I(\mathbf X;\mathbf Y).
\end{equation*}
Therefore, a trivial upper bound for the capacity is given by
\begin{equation}\label{1.02}
    C(u_p,u_a)\leq C_{G}=\frac{n}{2}\ln \left(1+\frac{\min (u_p,u_a)}{n}\right),
\end{equation}
where $C_G$ is achieved by a Gaussian input vector distributed as $N\left(\mathbf 0,\frac{\min(u_p,u_a)}{n}\mathbf{I}_n\right)$.

We formulate the optimization problem in the spherical domain. The rational behind this change of coordinates is due to the spherical symmetry of the white Gaussian noise and the constraints which, as it will be clear, enables us to perform the optimization problem only on the magnitude of the input. By writing the mutual information in terms of the differential entropies, we have
\begin{equation*}
    I(\mathbf{X};\mathbf{Y})=h(\mathbf{Y})-h(\mathbf{Y}|\mathbf{X})=h(\mathbf{Y})-\frac{n}{2}\ln2\pi e,
\end{equation*}
where the entropies are in nats. Motivated by the spherical symmetry of the white Gaussian noise and the constraints, $\mathbf Y$ and $\mathbf X$ can be written in spherical coordinates as
\begin{align*}
\mathbf Y=R\mathbf a(\mathbf \Psi)\ \ ,\ \ \mathbf X=P\mathbf a(\mathbf \Theta),
\end{align*}
where $R$ and $P$ denote the magnitude of the output and the input, respectively. $\mathbf \Psi=[\Psi_1,\Psi_2,\ldots,\Psi_{n-1}]^T$ and $\mathbf \Theta=[\Theta_1,\Theta_2,\ldots,\Theta_{n-1}]^T$ are, respectively, the phase vectors of the output and the input, in which $\Psi_i$, $\Theta_i\in [0,\pi] (i\in [1:n-2])$ and $\Psi_{n-1}$,$\Theta_{n-1}\in[0,2\pi).$ $\mathbf a(\mathbf \phi)=[a_1(\mathbf \phi),\ldots,a_n(\mathbf \phi)]^T$ is a unit vector in which
\begin{equation}
    a_k(\mathbf \phi)=\left\{\begin{array}{cc} \cos\phi_k\prod_{i=1}^{k-1}\sin\phi_i & k\in [1:n-1]\\ \prod_{i=1}^{k-1}\sin\phi_i & k=n \end{array}\right..\label{e4}
\end{equation}
As it will become clear later, this change of coordinates prevents the usage of the identity theorem for holomorphic functions of several complex variables. The optimization problem in (\ref{e2}) is equivalent to
\begin{equation}\label{e7}
    C(u_p,u_a)=\sup_{F_{P,\mathbf \Theta}(\rho,\mathbf \theta):P^2\leq u_p,\  E(P^2)\leq u_a}h(\mathbf Y)-\frac{n}{2}\ln 2\pi e.
\end{equation}
The differential entropy of the output is given by
\begin{align}
    h(\mathbf Y)&=-\int_{\mathbb{R}^n}f_{\mathbf Y}(\mathbf y)\ln f_{\mathbf Y}(\mathbf y)d\mathbf y \nonumber\\
    &=-\int_{0}^{\infty}\!\!\underbrace{\int_{0}^{\pi}\!\!\ldots\int_{0}^{\pi}}_{n-2\mbox{ times}}\!\!\int_{0}^{2\pi}f_{\mathbf Y}(\mathbf y(r,\mathbf \psi))\ln f_{\mathbf Y}(\mathbf y(r,\mathbf \psi))|\frac{\partial \mathbf y}{\partial(r,\mathbf \psi)}|d\mathbf\psi dr\nonumber\\
    &=-\int_{0}^{\infty}\!\!\underbrace{\int_{0}^{\pi}\!\!\ldots\int_{0}^{\pi}}_{n-2\mbox{ times}}\!\!\int_{0}^{2\pi}f_{R,\mathbf \Psi}(r,\mathbf \psi)\ln \frac{f_{R,\mathbf \Psi}(r,\mathbf \psi)}{|\frac{\partial \mathbf y}{\partial(r,\mathbf \psi)}|}d\mathbf\psi dr\nonumber\\
    &=h(R,\mathbf{\Psi})+\int_{0}^{\infty}f_R(r)\ln r^{n-1}dr\nonumber\\
    &\ \ \ +\sum_{i=1}^{n-2}\int_{0}^{\pi}f_{\Psi_i}(\psi_i)\ln\sin^{n-i-1}\psi_id\psi_i,\label{e6}
\end{align}
where $|\frac{\partial \mathbf y}{\partial(r,\mathbf \psi)}|( = r^{n-1}\prod_{i=1}^{n-2}\sin^{n-i-1}\psi_i)$ is the Jacobian of the transform.
The conditional pdf of $R,\mathbf \Psi$ conditioned on $P,\mathbf \Theta$ is given by
\begin{align}
   f_{R,\mathbf{\Psi}|P,\mathbf{\Theta}}(r,\mathbf{\psi}|\rho,\mathbf{\theta})&=\frac{1}{{(\sqrt{2\pi})}^n}e^{-\frac{r^2+\rho^2-2r\rho \mathbf a^T(\mathbf \theta)\mathbf a(\mathbf \psi)}{2}}r^{n-1}\nonumber\\
   &\ \ \ \times\prod_{i=1}^{n-2}\sin^{n-i-1}\psi_i.\label{e8}
\end{align}
From (\ref{e8}), the joint pdf of the magnitude and phases of the output is
\begin{equation}\label{e9}
    f_{R,\mathbf \Psi}(r,\mathbf \psi)=\int_{0}^{\infty}\!\!\!\!\!\underbrace{\int_{0}^{\pi}\!\!\ldots\int_{0}^{\pi}}_{n-2\mbox{ times}}\!\!\!\int_{0}^{2\pi}\!\!\!\!\!f_{R,\mathbf{\Psi}|P,\mathbf{\Theta}}(r,\mathbf{\psi}|\rho,\mathbf{\theta})
    d^nF_{P,\mathbf\Theta}(\rho,\mathbf\theta),
\end{equation}
in which $F_{P,\mathbf\Theta}(\rho,\mathbf\theta)$ denotes the joint CDF of $(P,\mathbf \Theta).$ By integrating (\ref{e9}) over the phase vector $\mathbf \psi$, we have
\begin{equation}\label{e10}
f_R(r)=\int_{0}^{\infty}L(r,\rho)f_P(\rho)d\rho,
\end{equation}
where \footnote{The reason that $L(r,\rho)$ is not a function of the phase vector $\mathbf \theta$ is due to the spherically symmetric distribution of the white Gaussian noise. In other words, $L(r,\rho)$ is the integral of the Gaussian pdf $N(\mathbf x,\mathbf I)$ over the surface of an n-sphere with radius $r$ which is invariant to the position of $\mathbf x$ as long as $\|\mathbf x\|=\rho$, i.e.
\begin{equation*}
   L(r,\|\mathbf x\|)=\int_{\|\mathbf y\|=r}\frac{e^{-\frac{\|\mathbf y- \mathbf x\|^2}{2}}}{(\sqrt{2\pi})^n}d\mathbf y= \frac{e^{-\frac{r^2+ \|\mathbf x\|^2}{2}}}{(\sqrt{2\pi})^n}\int_{\|\mathbf y\|=r}e^{\mathbf x^t\mathbf y}d\mathbf y
\end{equation*}
which is constant on $\|\mathbf x\|=\rho$. (\ref{e10}) implies that in the AWGN channel in (\ref{e1}), $f_R(r)$ is induced only by $f_P(\rho)$ and not $f_{\mathbf \Theta}(\mathbf \theta)$.}
\begin{equation*}
    L(r,\rho)=\underbrace{\int_{0}^{\pi}\ldots\int_{0}^{\pi}}_{n-2\mbox{ times}}\int_{0}^{2\pi}f_{R,\mathbf{\Psi}|P,\mathbf{\Theta}}(r,\mathbf{\psi}|\rho,\mathbf{\theta})d\psi_{n-1}\ldots d\psi_1.
\end{equation*}
It is obvious that
\begin{equation}\label{e13}
    h(R,\mathbf\Psi)\leq h(R)+\sum_{i=1}^{n-1}h(\Psi_i)\leq h(R)+\sum_{i=1}^{n-2}h(\Psi_i)+\ln2\pi,
\end{equation}
where the first inequality is tight iff the elements of $\{R,\Psi_1,\ldots,\Psi_{n-1}\}$ are mutually independent, and the second inequality becomes tight iff $\Psi_{n-1}$ is uniformly distributed over $[0,2\pi)$. From (\ref{e6}) and (\ref{e13}),
\begin{align}
    h(\mathbf Y)&\leq h(R)+\sum_{i=1}^{n-2}h(\Psi_i)+\int_{0}^{\infty}f_R(r)\ln r^{n-1}dr\nonumber\\
    &\ \ \ +\sum_{i=1}^{n-2}\int_{0}^{\pi}f_{\Psi_i}(\psi_i)\ln \sin^{n-i-1}\psi_id\psi_i+\ln2\pi.\label{e14}
\end{align}
For the sake of readability, the following change of variables is helpful
\begin{equation}\label{e15}
    V=\frac{R^n}{n}\ ,\ U_i=\int_{0}^{\Psi_i}\sin^{n-i-1}\delta d\delta\ ,\ i\in[1:n-2].
\end{equation}
Since $R\geq0$ and $\Psi_i\in [0,\pi](i\in[1:n-2])$, it is easy to show that the two mappings $R\to V$ and $\Psi_i\to U_i$ (defined in (\ref{e15})) are invertible. Also, the support set of $U_i$ is $S_{U_i}=[0,\alpha_i]$ where $\alpha_i = \frac{\sqrt{\pi}\Gamma(\frac{n-i}{2})}{\Gamma(\frac{n-i+1}{2})}$ (the Gamma function is defined as $\Gamma(t)=\int_{0}^{\infty}x^{t-1}e^{-x}dx$.)
From (\ref{e10}), the pdf of $V$ is \footnote{The existence of $f_V(v)$ is guaranteed by the Gaussian distribution of the additive noise.}
\begin{equation}\label{e16}
    f_V(v)=f_V(v;F_P)=\int_{0}^{\infty}K_n(v,\rho)dF_P(\rho),
\end{equation}
where the notation $;F_P$ in $f_V(v;F_P)$ is to emphasize that $V$ has been induced by $F_P$. Not that the integral transform in (\ref{e16}) is invertible as shown in Appendix \ref{a2}. The kernel $K_n(v,\rho)$ is given by
\begin{align}
    K_n(v,\rho)&=\frac{L(\sqrt[n]{nv},\rho)}{(\sqrt[n]{nv})^{n-1}}\nonumber\\&=\underbrace{\int_{0}^{\pi}\!\!\ldots\int_{0}^{\pi}}_{n-2\mbox{ times}}\!\!\int_{0}^{2\pi}\!\!\!\frac{1}{{(\sqrt{2\pi})}^n}e^{-\frac{(\sqrt[n]{nv})^2+\rho^2-2\sqrt[n]{nv}\rho \mathbf a^T(\mathbf \theta)\mathbf a(\mathbf \psi)}{2}}\nonumber\\&\ \ \ \ \ \ \ \ \ \ \ \ \ \ \ \ .\prod_{i=1}^{n-2}\sin^{n-i-1}\psi_id\psi_{n-1}\ldots d\psi_1\label{e16.5}\\
    &=e^{-\frac{(\sqrt[n]{nv})^2+\rho^2}{2}}\left\{\begin{array}{cc} \frac{I_{\frac{n}{2}-1}(\rho\sqrt[n]{nv})}{(\rho\sqrt[n]{nv})^{\frac{n}{2}-1}} & \rho v\neq0\\ \frac{1}{\Gamma(\frac{n}{2})2^{\frac{n}{2}-1}} & \rho v=0 \end{array}\right.,\forall n\geq2,\label{e17}
\end{align}
where $I_\alpha(.)$ is the modified bessel function of the first kind and order $\alpha$. The calculations are provided in Appendix \ref{a0}. Note that $K_n(v,\rho)$ is continuous on its domain.
The differential entropy of $V$ is
\begin{align}
    h(V)&=h(V;F_P)\nonumber\\
    &=-\int_{0}^{\infty}f_V(v;F_P)\ln f_V(v;F_P)dv\nonumber\\
    &=-\int_{0}^{\infty}f_R(r)\ln\frac{f_R(r)}{r^{n-1}}dr\label{e18}.
\end{align}
The differential entropy of $U_i$ is given by
\begin{align}
    h(U_i)&=-\int_{S_{U_i}}f_{U_i}(u)\ln f_{U_i}(u)du\nonumber\\
    &=-\int_{0}^{\pi}f_{\Psi_i}(\psi_i)\ln\frac{f_{\Psi_i}(\psi_i)}{\sin^{n-i-1}\psi_i}d\psi_i\ \ ,\ \ i\in[1:n-2].\label{e19}
\end{align}
Rewriting (\ref{e7}), we have
\begin{align}
    C(u_p,u_a)&=\sup_{F_{P,\mathbf\Theta}(\rho,\mathbf\theta):P^2\leq u_p,E[P^2]\leq u_a}h(\mathbf Y)-\frac{n}{2}\ln2\pi e \nonumber\\
    &\leq\!\!\!\sup_{F_{P,\mathbf\Theta}(\rho,\mathbf\theta):P^2\leq u_p,E[P^2]\leq u_a}\!\!\!h(V;F_P)+\sum_{i=1}^{n-2}h(U_i)\nonumber\\
    &\ \ \ +(1-\frac{n}{2})\ln2\pi-\frac{n}{2}\label{e21}\\
    &\leq\!\!\!\sup_{F_P(\rho):P^2\leq u_p,E[P^2]\leq u_a}\!\!\!h(V;F_P)+\sum_{i=1}^{n-2}\ln\alpha_i\nonumber\\
    &\ \ \ +(1-\frac{n}{2})\ln2\pi-\frac{n}{2}\label{e22},
\end{align}
where (\ref{e21}) results from (\ref{e14}), (\ref{e18}) and (\ref{e19}). (\ref{e22}) is due to the fact that since $S_{U_i}$ (the support of $U_i$) is bounded, $h(U_i)$ is maximized when $U_i$ is uniformly distributed. It is easy to verify that if the magnitude and phases of the input are mutually independent with the phases having the distributions as
\begin{equation}\label{e23}
    \Theta_{n-1}\sim U[0,2\pi)\ ,\ f_{\Theta_i}(\theta_i)=\alpha_i^{-1}\sin^{n-i-1}\theta_i\ \ ,\ \  i\in[1:n-2],
\end{equation}
the magnitude and phases of the output become mutually independent with the phases having the distributions as
\begin{equation}\label{e24}
    \Psi_{n-1}\sim U[0,2\pi)\ ,\ f_{\Psi_i}(\psi_i)=\alpha_i^{-1}\sin^{n-i-1}\psi_i\ \ ,\ \  i\in[1:n-2],
\end{equation}
where $\alpha_i = \frac{\sqrt{\pi}\Gamma(\frac{n-i}{2})}{\Gamma(\frac{n-i+1}{2})}$. In other words, having the input distribution
\begin{equation}\label{e25}
   F_{P,\mathbf \Theta}(\rho,\mathbf \theta)=\frac{\theta_{n-1}}{2\pi}F_P(\rho)\prod_{i=1}^{n-2}\int_{0}^{\theta_i}\alpha_i^{-1}\sin^{n-i-1}\theta d\theta
\end{equation}
results in
\begin{equation}\label{e26}
   F_{R,\mathbf \Psi}(r,\mathbf \psi)=\frac{\psi_{n-1}}{2\pi}F_R(r)\prod_{i=1}^{n-2}\int_{0}^{\psi_i}\alpha_i^{-1}\sin^{n-i-1}\psi d\psi.
\end{equation}
The above result can be easily checked either by solving for $f_{R,\mathbf \Psi}(r,\mathbf \psi)$ in (\ref{e9}) or by the fact that the summation of two independent spherically symmetric random vectors is still spherically symmetric.\footnote{The magnitude and the unit vector of a spherically symmetric random vector are independent and the unit vector is uniformly distributed on the unit ball. It can be verified that this property is equivalent to the vector having the distribution of (\ref{e26}) in spherical coordinates.}
Also, note that having $\Psi_i\ (i=1,\ldots,n-2)$ distributed as in (\ref{e24}) implies uniform $U_i$ on $[0,\alpha_i]\ (i=1,\ldots,n-2).$ It can be observed that the input pdf in (\ref{e25}) makes the inequalities in (\ref{e21}) and (\ref{e22}) tight. Since the constraint is only on the magnitude of the input and $f_V(v)$ is induced only by $f_{P}(\rho)$, it is concluded that the optimal input distribution must have mutually independent phases and magnitude with the phases being distributed as in (\ref{e23}). Therefore,
\begin{align}
    C(u_p,u_a)&=\sup_{F_P(\rho):P^2\leq u_p,E[P^2]\leq u_a}h(V;F_P)\nonumber\\
    &\ \ \ +\sum_{i=1}^{n-2}\ln\alpha_i+(1-\frac{n}{2})\ln2\pi-\frac{n}{2}.\label{e27}
\end{align}
Before proceeding further, it is interesting to check whether the problem in (\ref{e27}) boils down to the classical results when the peak power constraint is relaxed (i.e., $u_p\to \infty$). From the definition of $V$,
\begin{equation*}
    E[V^{\frac{2}{n}}]=\frac{1}{\sqrt[n]{n^2}}E[n+P^2].
\end{equation*}
This can be verified by a change of variable (i.e., $V=\frac{R^n}{n}$) and using the derivative of (\ref{e48}) (in Appendix \ref{a2}) with respect to $\beta$. Therefore, when $u_p\to\infty$, the problem in (\ref{e27}) becomes maximization of the differential entropy over all the distributions having a bounded moment of order $\frac{2}{n}$ which is addressed in Appendix \ref{a5} for an arbitrary moment. Substituting $m$ with $\frac{2}{n}$ and $A$ with $\frac{n+u_a}{\sqrt[n]{n^2}}$ in (\ref{e66}), the optimal distribution for $V$ is obatined and from (\ref{e16}), the corresponding $f_{P^*}(\rho)$ has the general Rayleigh distribution as
\begin{equation*}
    f_{P^*}(\rho)=\frac{n^{\frac{n}{2}}\rho^{n-1}e^{-\frac{n\rho^2}{2u_a}}}{2^{\frac{n-2}{2}}u_a^{\frac{n}{2}}\Gamma(\frac{n}{2})},
\end{equation*}
which is the only solution, since (\ref{e16}) is an invertible transform (see Appendix \ref{a2}). Furthermore, it can be verified that the maximum is
\begin{equation}\label{e27.3}
    C(\infty,u_a)=\frac{n}{2}\ln(1+\frac{u_a}{n}),
\end{equation}
which coincides with the classical results for the identity channel matrix \cite{Telatar}.

Similar to \cite{Smith} and \cite{Shamai}, we define the marginal entropy density of $V$ as
\begin{equation}\label{e28}
   \tilde{h}_V(x;F_P)=-\int_{0}^{\infty}K_n(v,x)\ln f_V(v;F_P)dv,
\end{equation}
which satisfies
\begin{equation*}
    h(V;F_P)=\int_{0}^{\infty}\tilde{h}_V(\rho;F_P)dF_P(\rho).
\end{equation*}
(\ref{e28}) is shown to be an invertible transform in Appendix \ref{a2} and this property will become useful later on.
\section{Main results}\label{s2.2}
Let $\epsilon_P$ denote the set of points of increase\footnote{A point $Z$ is said to be a point of increase of a distribution if for any open set $\Gamma$ containing $Z$, we have $\mbox{Pr}\{\Gamma\}>0.$} of $F_P(\rho)$ in the interval $[0,\sqrt{u_p}]$. The main result of the paper is given in the following theorem.

\textbf{Theorem.} The supremization in (\ref{e27}), which is for the identity channel matrix, has a unique solution and the optimal input achieving the supremum (and therefore the maximum) has the following distribution in the spherical coordinates,
\begin{equation}\label{e30}
   F^*_{P,\mathbf \Theta}(\rho,\mathbf \theta)=\frac{\theta_{n-1}}{2\pi}F_P^*(\rho)\prod_{i=1}^{n-2}\int_{0}^{\theta_i}\alpha_i^{-1}\sin^{n-i-1}\theta d\theta,
\end{equation}
where $F_P^*(\rho)$ has a finite number of points of increase (i.e., $\epsilon_{P^*}$ has a finite cardinality). Further, the necessary and sufficient condition for $F_{P}^*(\rho)$ to be optimal is the existence of a $\lambda(\geq0)$ for which
\begin{align}
    \tilde{h}_V(\rho;F_{P}^*)&\leq h(V;F_{P}^*)+\lambda(\rho^2-u_a)\ ,\ \forall\rho\in[0,\sqrt{u_p}]\label{e301}\\
    \tilde{h}_V(\rho;F_{P}^*)&= h(V;F_{P}^*)+\lambda(\rho^2-u_a)\ ,\ \forall\rho\in\epsilon_{P^*}.\label{e302}
\end{align}
Note that when the average power constraint is relaxed (i.e., $u_a\geq u_p$), $\lambda=0$.
\begin{proof}
The phases of the optimal input distribution have already been shown to be mutually independent and have the distribution in (\ref{e23}) being independent of the magnitude. Therefore, it is sufficient to show the optimal distribution of the input magnitude. This is proved by reductio ad absurdum. In other words, it is shown that having an infinite number of points of increase results in a contradiction. The detailed proof is given in Appendix \ref{app}.
\end{proof}
\textbf{Remark 1}. When the average power constraint is relaxed (i.e. $u_a\geq u_p$), the following input distribution is asymptotically ($\frac{u_p}{n}\to 0$) optimal
\begin{equation}\label{e59.15}
    F^{**}_{P,\mathbf \Theta}(\rho,\mathbf \theta)=\frac{\theta_{n-1}}{2\pi}u(\rho-\sqrt{u_p})\prod_{i=1}^{n-2}\int_{0}^{\theta_i}\alpha_i^{-1}\sin^{n-i-1}\theta d\theta,
\end{equation}
where $u(.)$ is the unit step function. Further, the resulting capacity is given by
\begin{equation*}
    C(u_p,u_p)\approx \frac{u_p}{2}\ \ \mbox{when}\ \ \frac{u_p}{n}\ll 1.
\end{equation*}
Later, in the numerical results section, we observe that the density in (\ref{e59.15}) remains optimal for the non-vanishing ratio $\frac{u_p}{n}$ when it is below a certain threshold.
\begin{proof}
Since the density in (\ref{e59.15}) has spherical symmetry, it is sufficient to show that $F^{**}_P(\rho)=u(\rho-\sqrt{u_p})$ is optimal when $\frac{u_p}{n}\to 0$. From (\ref{1.02}), we have
\begin{equation}\label{e59.14}
    \lim_{\frac{u_p}{n}\to 0}C(u_p,u_a)\leq \frac{u_p}{2}.
\end{equation}
The CDF $F^{**}_P(\rho)=u(\rho-\sqrt{u_p})$ induces the following output pdf
\begin{equation}
    f_V(v;F^{**}_P)=K_n(v,\sqrt{u_p})=e^{-\frac{(\sqrt[n]{nv})^2+u_p}{2}}\frac{I_{\frac{n}{2}-1}(\sqrt{u_p}\sqrt[n]{nv})}{(\sqrt{u_p}\sqrt[n]{nv})^{\frac{n}{2}-1}}.\label{e59.16}
\end{equation}

\begin{figure*}[!t]
\normalsize
\begin{align}
    \lim_{\frac{u_p}{n}\to 0} h(V;F^{**}_P)&=\lim_{\frac{u_p}{n}\to 0}-\int_{0}^{\infty}f_V(v;F^{**}_P)\ln f_V(v;F^{**}_P)dv\nonumber\\
    &=\lim_{\frac{u_p}{n}\to 0}\int_{0}^{\infty}e^{-\frac{(\sqrt[n]{nv})^2+u_p}{2}}\frac{I_{\frac{n}{2}-1}(\sqrt{u_p}\sqrt[n]{nv})}{(\sqrt{u_p}\sqrt[n]{nv})^{\frac{n}{2}-1}}\left[\frac{(\sqrt[n]{nv})^2+u_p}{2}-\ln \left(\frac{I_{\frac{n}{2}-1}(\sqrt{u_p}\sqrt[n]{nv})}{(\sqrt{u_p}\sqrt[n]{nv})^{\frac{n}{2}-1}}\right)\right]dv\nonumber\\
    &=\frac{n}{2}+\ln \left(\Gamma(\frac{n}{2})2^{\frac{n}{2}-1}\right)\nonumber\\
    &\ \ \ +\lim_{\frac{u_p}{n}\to 0}\left\{u_p-\int_{0}^{\infty}e^{-\frac{(\sqrt[n]{nv})^2+u_p}{2}}\frac{I_{\frac{n}{2}-1}(\sqrt{u_p}\sqrt[n]{nv})}{(\sqrt{u_p}\sqrt[n]{nv})^{\frac{n}{2}-1}}\ln \left(1+\frac{u_p(\sqrt[n]{nv})^2}{2n}\right)dv\right\}\label{eapp}\\
    &=\frac{n}{2}+\ln \left(\Gamma(\frac{n}{2})2^{\frac{n}{2}-1}\right)+\lim_{\frac{u_p}{n}\to 0}\left\{u_p - \frac{u_p}{n}(\frac{n+u_p}{2})\right\}\label{eapp2}\\
    &=\frac{n}{2}+\ln \left(\Gamma(\frac{n}{2})2^{\frac{n}{2}-1}\right)+\frac{u_p}{2}.\label{eapp3}
\end{align}
\hrulefill
\vspace*{4pt}
\end{figure*}

When $\frac{u_p}{n}$ is small, the entropy of $V$ is given by (\ref{eapp3}) on top of the next page. In (\ref{eapp}), we have approximated the modified bessel function with the first two terms in its power series expansion as follows
\begin{equation*}
    I_n(x)\approx\frac{x^n}{\Gamma(n+1)2^n}(1+\frac{x^2}{4(n+1)})\ \ ,\ \ \frac{x}{n}\to 0.
\end{equation*}
In (\ref{eapp2}), we use the approximation $\ln(1+x)\approx x\ (x\ll1)$ and in (\ref{eapp3}), the higher order term is neglected. Given the input distribution $F^{**}_P$, the achievable rate with small ratio $\frac{u_p}{n}$ is given by (see (\ref{e27}))
\begin{equation}\label{e59.20}
    \lim_{\frac{u_p}{n}\to 0} h(V;F^{**}_P)+\sum_{i=1}^{n-2}\ln\alpha_i+(1-\frac{n}{2})\ln2\pi-\frac{n}{2}=\frac{u_p}{2},
\end{equation}
where we have used the fact that
\begin{equation*}
    \sum_{i=1}^{n-2}\ln\alpha_i=-\ln \Gamma(\frac{n}{2}) + \frac{n-2}{2}\ln \pi.
\end{equation*}
From (\ref{e59.20}) and (\ref{e59.14}), it is concluded that the pdf in (\ref{e59.15}) is asymptotically optimal for $\frac{u_p}{n}\ll1$ when $u_p\leq u_a$. Note that the distribution in (\ref{e59.15}) is not the only asymptotically optimal distribution. There are many possible alternatives, one of which, for example, is the binary PAM in each dimension with the points $-\sqrt{\frac{u_p}{n}}$ and $\sqrt{\frac{u_p}{n}}$ which can be verified to have an achievable rate of $\frac{u_p}{2}$ when $\frac{u_p}{n}\ll 1$. Specifically, in the low peak power regime ($u_p\ll 1$), a sufficient condition for the input distribution to be asymptotically optimal is as follows. First, it has a constant magnitude at $\sqrt{u_p}$. Second, its $\Theta_1$ is independent of $(P,\Theta_2,\ldots,\Theta_{n-1})$ and has a zero first Fourier coefficient i.e.,
\begin{equation}\label{e26.5}
    \int_{0}^{\pi}e^{j\theta}f_{\Theta_1}(\theta)d\theta=0.
\end{equation}
The claim is justified by noting that fulfilling the second condition results in the spherical symmetric output distribution of (\ref{e26}) as follows. Using the approximation $e^x\approx1+x\ (x\ll1)$, at small values of $u_p$, (\ref{e8}) can be approximated as
\begin{align}
    f_{R,\mathbf{\Psi}|P,\mathbf{\Theta}}(r,\mathbf{\psi}|\rho,\mathbf{\theta})&\approx\frac{1}{{(\sqrt{2\pi})}^n}e^{-\frac{r^2+\rho^2}{2}}(1+r\rho \mathbf a^T(\mathbf \theta)\mathbf a(\mathbf \psi))\nonumber\\
    &\ \ \ \times r^{n-1}\prod_{i=1}^{n-2}\sin^{n-i-1}\psi_i.\label{e60}
\end{align}
If $\Theta_1$ is independent of $(\Theta_2,\ldots,\Theta_{n-1},P)$, substituting (\ref{e60}) in (\ref{e9}) results in
\begin{align}
    f_{R,\mathbf \Psi}(r,\mathbf \psi)&\approx\int_{0}^{\infty}\!\underbrace{\int_{0}^{\pi}\!\!\ldots\int_{0}^{\pi}\!}_{n-3\mbox{ times}}\int_{0}^{2\pi}\!\!\int_0^{\pi}\!\!\frac{1}{{(\sqrt{2\pi})}^n}e^{-\frac{r^2+\rho^2}{2}}r^{n-1}\nonumber\\
    &\ \ \ \times(1+r\rho \mathbf a^T(\mathbf \theta)\mathbf a(\mathbf \psi))\prod_{i=1}^{n-2}\sin^{n-i-1}\psi_i\nonumber\\
    &\ \ \ \ \ dF_{\Theta_1}(\theta_1)d^{n-1}F_{P,\mathbf\Theta_2^{n-1}}(\rho,\mathbf\theta_2^{n-1})\label{e61}
\end{align}
where $\mathbf\theta_2^{n-1}=(\theta_2,\theta_3,\ldots,\theta_{n-1})$. If $\Theta_1$ has a zero first Fourier coefficient, due to the structure of $\mathbf a(\mathbf \theta)$ (see (\ref{e4})), we have
\begin{equation*}
\int_{0}^{\pi}\mathbf a^T(\mathbf \theta)\mathbf a(\mathbf \psi)dF_{\Theta_1}(\theta_1)=0.
\end{equation*}
Therefore, (\ref{e61}) simplifies as
\begin{equation*}
    f_{R,\mathbf \Psi}(r,\mathbf \psi)\approx\int_{0}^{\infty}\frac{1}{{(\sqrt{2\pi})}^n}e^{-\frac{r^2+\rho^2}{2}}r^{n-1}\prod_{i=1}^{n-2}\sin^{n-i-1}\psi_idF_P(\rho)
\end{equation*}
which implies that when $u_p\to0$, having $\Theta_1$ independent of all other spherical variables with a zero first Fourier coefficient results in the output distribution in (\ref{e26}) which makes the inequalities (\ref{e21}) and (\ref{e22}) tight. Finally, fulfilling the first condition (i.e., having a constant magnitude at $\sqrt{u_p}$) validates the previous reasoning starting from (\ref{e59.16}).

The asymptotic optimality of the constant-magnitude signaling in (\ref{e59.15}) can alternatively be proved by inspecting the behavior of the marginal entropy density $\tilde{h}_V(\rho;F_P)$ when $\frac{u_p}{n}$ is sufficiently small. From (\ref{e16})
\begin{equation*}
    f_V(v;F_P)\to\frac{e^{-\frac{(\sqrt[n]{nv})^2}{2}}}{\Gamma(\frac{n}{2})2^{\frac{n}{2}-1}}\underbrace{\int_{0}^{\infty}e^{-\frac{\rho^2}{2}}dF_P(\rho)}_{\mbox{constant\ }=\ C}\ \mbox{when}\ \frac{u_p}{n}\to 0.
\end{equation*}
Therefore,
\begin{align}
    \tilde{h}_V(\rho;F_P)&=-\int_{0}^{\infty}e^{-\frac{(\sqrt[n]{nv})^2+\rho^2}{2}}\frac{I_{\frac{n}{2}-1}(\rho\sqrt[n]{nv})}{(\rho\sqrt[n]{nv})^{\frac{n}{2}-1}}\ln f_V(v;F_P)dv\nonumber\\
    &\to \int_{0}^{\infty}e^{-\frac{(\sqrt[n]{nv})^2+\rho^2}{2}}\frac{I_{\frac{n}{2}-1}(\rho\sqrt[n]{nv})}{(\rho\sqrt[n]{nv})^{\frac{n}{2}-1}}\left[\frac{(\sqrt[n]{nv})^2}{2}\right.\nonumber\\
    &\ \ \ \ \ \ \left.+\ln\left(\frac{\Gamma(\frac{n}{2})2^{\frac{n}{2}-1}}{C}\right)\right]dv\nonumber\\
    &= \frac{\rho^2+n}{2}+ \ln\left(\frac{\Gamma(\frac{n}{2})2^{\frac{n}{2}-1}}{C}\right)\label{e59.43}
\end{align}
It is obvious that (\ref{e59.43}) is a (strictly) convex (strictly) increasing function. Hence, the necessary and sufficient conditions in (\ref{e301}) and (\ref{e302}) are satisfied if and only if the input has only one point of increase at $\sqrt{u_p}$ which proves the asymptotic optimality of (\ref{e59.15}) for $\frac{u_p}{n}\ll 1$ and $u_a\geq u_p$.
\end{proof}

\textbf{Remark 2}. For a fixed SNR, the gap between Shannon capacity and the constant amplitude signaling decreases as $O(\frac{1}{n})$ for large values of $n$.
\begin{proof}
By writing the first two terms of the Taylor series expansion of the logarithm (i.e., $\ln(1+x)\approx x-\frac{x^2}{2}, x\ll1$), we have
\begin{equation*}
    \mbox{when }n\to\infty\ \ ,\ \ \frac{n}{2}\ln(1+\frac{u_p}{n})\approx \frac{u_p}{2}-\frac{u_p^2}{4n}.
\end{equation*}
From (\ref{eapp2}), the achievable rate obtained by the constant envelope signaling is
\begin{equation*}
    \mbox{when }n\to\infty\ \ ,\ \ I(\mathbf X;\mathbf Y)\approx \frac{u_p}{2}-\frac{u_p^2}{2n}.
\end{equation*}
This shows that the gap between achievable rate and the Shannon capacity decreases as $\frac{u_p^2}{4n}$($=O(\frac{1}{n})$), when $n$ goes to infinity.
\end{proof}
While remark 2 shows an asymptotic behavior of the gap, the following remark provides an analytical lower bound for any values of $n$.

\textbf{Remark 3}. The following lower bound holds for the capacity of constant amplitude signaling.
\begin{equation}\label{qw1}
    \sup_{F_{\mathbf X}(\mathbf x):\|\mathbf X\|^2=u_p}\!\!\!\!\!\!\!\!\!\!\!I(\mathbf X; \mathbf Y)\geq \frac{n-1}{2}\log\left(1+\frac{2^{\frac{2}{n-1}-1}u_p}{e\left[(n-1)\Gamma(\frac{n-1}{2})\right]^{\frac{2}{n-1}}}\right)
\end{equation}
\begin{proof}
Let $\mathbf{X}'$ and $\mathbf{X}'$ be defined as
\begin{equation}\label{qw2}
    \mathbf{X}'=\left[X_1,X_2,\ldots,X_{N-1},0\right]^T\ \ ,\ \ \mathbf{Y}'=\left[Y_1,Y_2,\ldots,Y_{N-1},0\right]^T.
\end{equation}
Due to the Markov chain $\mathbf{X}'\leftrightarrow\mathbf{X}\leftrightarrow\mathbf{Y}\leftrightarrow\mathbf{Y}'$ and the fact that $\|\mathbf X\|^2=u_p$ implies $\|\mathbf{X}'\|^2\leq u_p$, we can write
\begin{align}
    \sup_{\substack{F_{\mathbf X}(\mathbf x)\\\|\mathbf X\|^2=u_p}}\!\!\!\!\! I(\mathbf X;\mathbf Y)&\geq \sup_{F_{\mathbf{X}'}(\mathbf{x}'):\|\mathbf{X}'\|^2\leq u_p} \!\!\!\!\!\!\!\!\!\! I(\mathbf{X}';\mathbf{Y}')\nonumber\\
    &=\sup_{F_{\mathbf{X}'}(\mathbf{x}'):\|\mathbf{X}'\|^2\leq u_p} h(\mathbf{Y}';F_{\mathbf{X}'})\nonumber\\&\ \ \ -\frac{n-1}{2}\log 2\pi e\nonumber\\
    &\geq\!\!\! \sup_{F_{\mathbf{X}'}(\mathbf{x}'):\|\mathbf{X}'\|^2\leq u_p}\!\!\!\!\!\!\frac{n-1}{2}\log\left(2^{\frac{2}{n-1}h(\mathbf{X}')}+2\pi e\right)\nonumber\\&\ \ \ -\frac{n-1}{2}\log 2\pi e\label{epi}\\
    &=\frac{n-1}{2}\log\left(1+\frac{2^{\frac{2}{n-1}-1}u_p}{e\left[(n-1)\Gamma(\frac{n-1}{2})\right]^{\frac{2}{n-1}}}\right)\label{es}
\end{align}
where in (\ref{epi}), the $(n-1)$-dimensional EPI has been used\footnote{Note that the reduction of dimensions from $n$ to $n-1$ in (\ref{qw2}) is necessary. The reason is that the usage of the n-dimensional EPI is not permissible for the constant amplitude vector, since an $n$-dimensional vector with a fixed norm has at most $(n-1)$ degrees of freedom (or equivalently at most $(n-1)$-dimensional support).}
 and (\ref{es}) is due to the fact that for the $(n-1)$-dimensional vector $\mathbf{X}'$, we can write
\begin{equation}\label{qw3}
    \sup_{F_{\mathbf{X}'}(\mathbf{x}'):\|\mathbf{X}'\|^2\leq u_p}h(\mathbf{X}')=\log\left(\frac{2(\pi u_p)^\frac{n-1}{2}}{(n-1)\Gamma(\frac{n-1}{2})}\right)\ \ ,\ \ n\geq 2
\end{equation}
whose proof follows the same steps from (\ref{qw10}) to (\ref{e.8}) with $\lambda=0$ and $a = \frac{n}{(\sqrt{u_p})^n}$.
\end{proof}
The asymptotic decrease of the gap in remark 2 can be alternatively proved by using the lower bound in (\ref{es}) which is provided in Appendix \ref{altp}.

\textbf{Remark 4}. When $u_a < u_p$, the following input distribution is asymptotically ($u_a\to 0$) optimal
\begin{align}
    F^{**}_{P,\mathbf \Theta}(\rho,\mathbf \theta)&=\left[(1-\frac{u_a}{u_p})u(\rho)+\frac{u_a}{u_p}u(\rho-\sqrt{u_p})\right]\nonumber\\&\ \ \ \times\frac{\theta_{n-1}}{2\pi}\prod_{i=1}^{n-2}\int_{0}^{\theta_i}\alpha_i^{-1}\sin^{n-i-1}\theta d\theta\label{e59.51}
\end{align}
and the resulting capacity is given by
\begin{equation*}
    C(u_p,u_p)\approx \frac{u_a}{2}\ \ \mbox{when}\ \ u_a\ll 1.
\end{equation*}
\begin{proof}
The proof is given in Appendix \ref{app1}.
\end{proof}
Remarks 1 and 4 are essential for the initial stage of the simulation results when either $u_a$ or $u_p$ are assumed to be very small at first and afterwards they are increased gradually by a step size.

\textbf{Remark 5}. The fact that the magnitude of the optimal input distribution has a finite number of mass points remains unchanged if the average constraint in (\ref{e2}) is generalized as
\begin{equation}\label{e59.01}
    E(g(P))\leq u_a
\end{equation}
in which $g(z)$ is holomorphic on an open subset $\mathbb{D}(\subseteq\mathbb{C})$ which includes the non-negative real line (i.e., $\mathbb{R}_{\geq 0}\subset \mathbb{D}$).
\begin{proof}
The proof is given in Appendix \ref{pr4}.
\end{proof}
\textbf{Remark 6}. The peak power constraint in (\ref{e2}) can be generalized to
\begin{equation*}
    \|P\|^2\stackrel{a.s.}\in \mathbb{D}_{u_p}\subseteq [0,u_p].
\end{equation*}
\begin{proof} Since all the conditions (compactness, continuity, etc.) remain unchanged, the support of the optimal input distribution will be some concentric shells having the mass points of the magnitude in $\mathbb{D}_{u_p}$.
\end{proof}
\section{the mimo case with deterministic channel}\label{s3}
First, we consider the multiple-input single-output (MISO) channel in which (\ref{e1}) changes to
\begin{equation}\label{ee1}
    Y(t)=\mathbf{h}^T\mathbf{X}(t)+W(t)
\end{equation}
where $\mathbf{h}(\in \mathbb{R}^{n\times1})$ is the deterministic channel vector and $W\sim N(0,1).$
Let $X_{new}=\mathbf h^T\mathbf{X}$. The capacity of this channel under the peak and average power constraints is given by
\begin{align}
    C(u_p,u_a)&=\sup_{\substack{F_{\mathbf{X}}(\mathbf x):\|\mathbf X\|^2\leq u_p,\\E[\|\mathbf X\|^2]\leq u_a}}\!\!\!\!\!\!\!\!\!I(\mathbf X; Y)\nonumber\\
    &=\sup_{\substack{F_{\mathbf{X}}(\mathbf x):\|\mathbf X\|^2\leq u_p,\\E[\|\mathbf X\|^2]\leq u_a}}\!\!\!\!\!\!\!\!\!I(\mathbf X,X_{new}; Y)\label{ee2}\\
    &=\sup_{\substack{F_{\mathbf{X}}(\mathbf x):\|\mathbf X\|^2\leq u_p,\\E[\|\mathbf X\|^2]\leq u_a}}\!\!\!\!\!\!\!\!\!I(X_{new}; Y)+\underbrace{I(\mathbf{X}; Y|X_{new})}_{=0}\label{ee3}\\
    &=\sup_{\substack{F_{\mathbf{X}}(\mathbf x):\|\mathbf X\|^2\leq u_p,\\E[\|\mathbf X\|^2]\leq u_a}}\!\!\!\!\!\!\!\!\!I(X_{new}; Y)\nonumber\\
    &\leq \sup_{\substack{F_{X_{new}}(x):|X_{new}|\leq \sqrt{u_p}\|\mathbf{h}\|,\\ E[|X_{new}|^2]\leq u_a\|\mathbf h\|^2}}\!\!\!\!\!\!\!\!\!I(X_{new}; Y)\label{ee4}
\end{align}
where (\ref{ee2}) is due to the fact that $X_{new}$ is a function of $\mathbf{X}$ and (\ref{ee3}) is a result of the following Markov chain $\mathbf{X}\longrightarrow X_{new}\longrightarrow Y$. (\ref{ee4}) is due to the fact that any input cdf having the support $\|\mathbf X\|^2\leq u_p$ and satisfying $E[\|\mathbf X\|^2]\leq u_a$ induces a cdf for $X_{new}$ with the support in $[-\sqrt{u_p}\|\mathbf{h}\|,\sqrt{u_p}\|\mathbf{h}\|]$ and satisfying $E[|X_{new}|^2]\leq u_a\|\mathbf h\|^2$. This could be readily verified by the following convex optimization problem
\begin{align*}
  &\ \ \ \max_{\mathbf x}{\mathbf{h}^T\mathbf{x}}\nonumber\\
  & \mbox{S.t. } \|\mathbf x\|^2\leq u_p
\end{align*}
where the maximum is $\sqrt{u_p}\|\mathbf{h}\|$ and it is achieved when $\mathbf x$ is matched to the channel (i.e., $\mathbf x=\sqrt{u_p}\frac{\mathbf h}{\|\mathbf{h}\|}$). Further, from Cauchy-Shwartz inequality, we have
\begin{equation}\label{ee5}
    E[|X_{new}|^2]=E[|\mathbf h^t\mathbf{X}|^2]\leq E[\|\mathbf h\|^2]E[\|\mathbf X\|^2]\leq u_a\|\mathbf h\|^2
\end{equation}
where the inequalities change to equality iff $\mathbf X$ is in the direction of $\mathbf h$ and $E[\|\mathbf X\|^2]=u_a$.

The supremization in (\ref{ee4}) is the same problem of finding the capacity of a scalar Gaussian channel which has been addressed in \cite{Smith} where it was shown that the optimal input distribution is a pmf over a finite set of points in the interval defined by the peak power constraint and also it satisfies the average power inequality with equality. It is obvious that having $\mathbf{X}$ located on the hyperplane $\mathbf h^T\mathbf{X}=e_i$ (confined in the ball $\|\mathbf X\|^2\leq u_p$) with probability $p_i$ results in having $X_{new}$ equal to the mass point $e_i\in[-\sqrt{u_p}\|\mathbf{h}\|,\sqrt{u_p}\|\mathbf{h}\|]$ with probability $p_i$. If the average power constraint is relaxed (i.e., $u_a\geq u_p$), the support of the capacity-achieving distribution of the MISO channel with the input bounded in a ball becomes a finite number of hyper planes confined in that ball (all of these hyperplane have the normal vector $\mathbf{h}$). Note that the discrete amplitude property is no longer a necessity for the optimal input distribution in contrast to the MIMO with identity channel. In other words, the necessary and sufficient condition for the optimality is that $\mathbf X$ is located on  each of these hyperplanes with the corresponding probabilities. There is a common characteristic of the optimal input distribution in both the MIMO (with identity channel) and MISO scenarios which is the fact that the support of the optimal input distribution does not include any open set in $\mathbb{R}^n$. Finally, if the average power constraint is active (i.e., $u_a< u_p$), the support of the optimal input becomes a finite number of mass points in the direction of $\mathbf h$ (from (\ref{ee5}) and the fact that $E[|X_{new}|^2]=u_a\|\mathbf h\|^2$) and confined in the ball $\|\mathbf X\|^2\leq u_p$.

For the general deterministic MIMO channel, we have
\begin{equation}\label{e30.01}
    \mathbf{Y'}(t)=\mathbf H\mathbf{X'}(t)+\mathbf{W'}(t)
\end{equation}
where $\mathbf H\in\mathbb{R}^{n_r\times n_t}$ denotes the deterministic channel. By an SVD (i.e., $\mathbf H=\mathbf D\mathbf \Lambda \mathbf N^T$ where $\mathbf D\in\mathbb{R}^{n_r\times n_r}$, $\mathbf \Lambda\in\mathbb{R}^{n_r\times n_t}$, $\mathbf N\in\mathbb{R}^{n_t\times n_t}$), we get
\begin{equation}\label{e30.02}
    \tilde{\mathbf{Y'}}(t)=\mathbf D^T\mathbf{Y'}(t)=\mathbf \Lambda\underbrace{\mathbf N^T\mathbf{X'}(t)}_{\tilde{\mathbf{X'}}(t)}+\underbrace{\mathbf D^T\mathbf{W'}(t)}_{\tilde{\mathbf{W'}}(t)}.
\end{equation}
Let $n=\mbox{rank}(\mathbf H)$ and $\mathbf{Q}(t)$ be the first $n$ elements of the vector $\tilde{\mathbf{Q'}}(t)$ (for $\mathbf{Q}=\mathbf{Y}, \mathbf{X}\mbox{ and }\mathbf{W}$). It is obvious that (\ref{e30.02}) is equivalent to the following
\begin{equation}\label{e30.03}
    \mathbf{Y}(t)=\mathbf{X}(t)+\mathbf{N}(t)
\end{equation}
with the noise distributed as $N(\mathbf{0},\mathbf\Sigma)$ where $\mathbf\Sigma=\mbox{diag}\{\lambda_1^{-2},\lambda_2^{-2},\ldots,\lambda_n^{-2}\}$ and $\lambda_i$ ($i\in[1:n]$) is the $i^{th}$ singular value of $\mathbf H$. Therefore, the capacity of the deterministic channel in (\ref{e30.01}) is the same as the capacity of the additive non-white Gaussian noise channel in (\ref{e30.03}). It is assumed that the condition number of $\mathbf H$ is not unity, since in that case, it becomes equivalent to the scenario with identity channel matrix discussed in section \ref{s2.1}. From now on, we consider $n=2$.

\begin{figure*}[!t]
  \centering
  \begin{minipage}[c]{1\textwidth}
  \centering
  \subfigure[Elliptical Constraints]{

    \includegraphics[width=.4\textwidth]{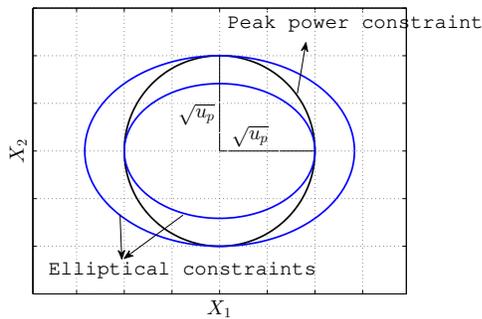}

  }\label{yaza}%
  \vfill
  \subfigure[Cubic Constraints]{
    \includegraphics[width=.4\textwidth]{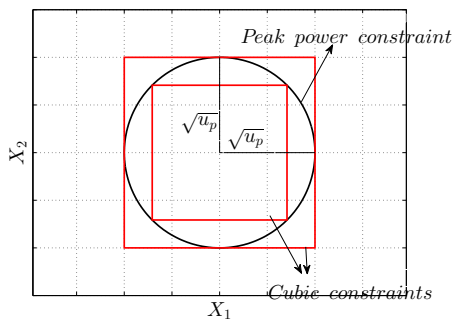}

  }%
  \vfill
  \subfigure[Modified Cubic Constraint]{
    \includegraphics[width=.4\textwidth]{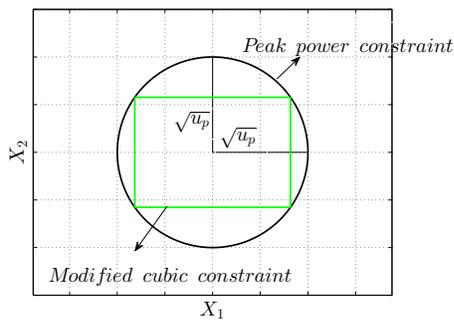}

  }
   \caption{Weakening or strengthening the peak power constraint for $n=2$ and $\lambda_1^2=2\lambda_2^2$.}\label{yaz}
\end{minipage}
\end{figure*}

Two possible changes of coordinates are as follows. Motivated by the elliptical symmetry of the noise, $\mathbf X$ and $\mathbf Y$ could be written in the following elliptical coordinates
\begin{equation}\label{e30.1}
    \mathbf Y=R\mathbf\Sigma^{\frac{1}{2}}\mathbf a(\mathbf \Psi)\ \ ,\ \ \mathbf X=P\mathbf\Sigma^{\frac{1}{2}}\mathbf a(\mathbf \Theta)
\end{equation}
and using a similar approach as in section \ref{s2.1}, the optimization problem becomes
\begin{equation}\label{e30.11}
    C(u_p,u_a)=\sup_{\substack{F_{P,\Theta(\rho,\theta)}:P^2\mathbf a^T(\theta)\mathbf\Sigma\mathbf a(\theta)\leq u_p, \\E[P^2\mathbf a^T(\theta)\mathbf\Sigma\mathbf a(\theta)]\leq u_a}}\!\!\!\!\!\!\!\!\!\!\!\!\!\!\!\!\!\!\!\!\!\!h(V,\Psi;F_{P,\Theta})-\ln 2\pi e,
\end{equation}
where $V=\frac{R^2}{2}.$ The joint entropy of the output variables is given by
\begin{equation}\label{e30.06}
    h(V,\Psi;F_{P,\Theta})=\int_{0}^{\infty}\int_{0}^{2\pi}\tilde{h}_{V,\Psi}(\rho,\theta;F_{P,\Theta})d^2F_{P,\Theta}(\rho,\theta),
\end{equation}
where the joint marginal entropy density writes as
\begin{align}
    \tilde{h}_{V,\Psi}(\rho,\theta;F_{P,\Theta})&=-\int_{0}^{\infty}\!\!\!\int_{0}^{2\pi}K(v,\psi,\rho,\theta)\nonumber\\&\ \ \ \ \ \ \ \ \ \times\ln f_{V,\Psi}(v,\psi;F_{P,\Theta})d\psi dv\label{e30.07},
\end{align}
in which
\begin{equation}\label{e30.08}
    f_{V,\Psi}(v,\psi;F_{P,\Theta})=\int_{0}^{\infty}\int_{0}^{2\pi}K(v,\psi,\rho,\theta)d^2F_{P,\Theta}(\rho,\theta),
\end{equation}
and
\begin{equation}\label{e30.12}
    K(v,\psi,\rho,\theta)=\frac{1}{2\pi}e^{-v-\frac{\rho^2}{2}+\rho\sqrt{2v}\cos(\psi-\theta)}.
\end{equation}
Alternatively, due to the spherical symmetry of the constraint, the input and the output could be written in the spherical coordinates in which
\begin{equation}\label{e30.05}
    C(u_p,u_a)=\sup_{F_{P,\Theta(\rho,\theta)}:P^2\leq u_p,E[P^2]\leq u_a}\!\!\!\!\!\!\!\!\!\!\!\!\!\!\!\!\!\!h(V,\Psi;F_{P,\Theta})-\ln(2\pi e\sqrt{|\mathbf\Sigma|}).
\end{equation}
(\ref{e30.06}) to (\ref{e30.08}) remain unchanged, while the kernel is given by
\begin{equation}\label{e30.09}
    K(v,\psi,\rho,\theta)=\frac{e^{-\frac{1}{2}\left[\sqrt{2v}\mathbf a(\psi)-\rho\mathbf a(\theta)\right]^T\mathbf\Sigma^{-1}\left[\sqrt{2v}\mathbf a(\psi)-\rho\mathbf a(\theta)\right]}}{2\pi\sqrt{|\mathbf\Sigma|}}.
\end{equation}
Using neither of the above coordinates makes the separation of the magnitude and the phases possible as done in (\ref{e13}). This is due to the different symmetries of the noise (elliptical) and the peak power constraint (spherical). Since the conditions of compactness, convexity and continuity remain unchanged, we can only proceed up to the point of writing the necessary and sufficient conditions for the joint cdf $F_{P,\Theta}(\rho,\theta)$ to be the optimal solution. By using the spherical coordinates, the necessary and sufficient conditions for the optimal input distribution is given by
\begin{align}
    \tilde{h}_{V,\Psi}(\rho,\theta;F_{{P,\Theta}}^*)&\leq h(V,\Psi;F_{{P,\Theta}}^*)+\lambda(\rho^2-u_a)\nonumber\\&\ \ \ \ ,\forall \rho\in[0,\sqrt{u_p}],\forall \theta\in [0,2\pi)\label{e30.15},\\
    \tilde{h}_{V,\Psi}(\rho,\theta;F_{{P,\Theta}}^*)&= h(V,\Psi;F_{{P,\Theta}}^*)+\lambda(\rho^2-u_a)\nonumber\\&\ \ \ \ ,\ \forall(\rho,\theta)\in\epsilon_{{P,\Theta}}^*,\label{e30.16}
\end{align}
where $\epsilon_{{P,\Theta}}^*$ is the set of points of increase in $F_{{P,\Theta}}^*$. 
To make the problem caused by the different symmetries of the noise and the constraint more clear, let's assume $\lambda_1 = \lambda_2$ (i.e., as in the previous section with identity channel.) In this case, we rewrite the optimization problem as
\begin{equation}\label{e30.17}
    C(u_p,u_a)=\sup_{F_{P,\Theta(\rho,\mathbf\theta)}:P^2\leq u_p,\ E[P^2]\leq u_a}\!\!\!\!\!\!h(V,\Psi;F_{P,\Theta})-\ln(2\pi e\lambda_1^2).
\end{equation}
It is already known that the optimal solution must have independent phase and magnitude with the former being uniformly distributed on $[0,2\pi).$ This can alternatively be inferred from the above necessary and sufficient conditions as follows. Let $f^*_{P,\Theta}(\rho,\theta)=f^*_P(\rho)f^*_{\Theta|P}(\theta|\rho)$ denote the (unique) solution of (\ref{e30.17}) with $\epsilon_{{P,\Theta}}^*$ as its points of increase. Let the pdf $l^{\epsilon}_{P,\Theta}$ be defined as
\begin{equation*}
    l^{\epsilon}_{P,\Theta}(\rho,\theta)=f^*_P(\rho)f^*_{\Theta|P}(\theta-\epsilon|\rho),
\end{equation*}
where $\epsilon$ is a constant arbitrarily chosen from $(0,2\pi)$. Let $L^{\epsilon}_{P,\Theta}$ be the corresponding CDF. It can be easily verified that
\begin{equation}\label{e30.20}
 f_{V,\Psi}(v,\psi;L^{\epsilon}_{P,\Theta})=f_{V,\Psi}(v,\psi-\epsilon;F^*_{P,\Theta})
\end{equation}
and therefore,
\begin{equation*}
   h(V,\Psi;L^{\epsilon}_{P,\Theta})=h(V,\Psi;F^*_{P,\Theta}).
\end{equation*}
Since $L^{\epsilon}_{P,\Theta}$ satisfies the constraints and the optimal solution is unique, it is concluded that
\begin{equation*}
    f^*_{P,\Theta}(\rho,\theta)=l^{\epsilon}_{P,\Theta}(\rho,\theta)
\end{equation*}
which in turn results in
\begin{equation*}
    f_{\Theta|P}(\theta|\rho)=f_{\Theta|P}(\theta-\epsilon|\rho).
\end{equation*}
Since $\epsilon\in(0,2\pi)$ was chosen arbitrarily, we conclude that $f_{\Theta|P}(\theta|\rho)=f_{\Theta}(\theta)=\frac{1}{2\pi}.$ The problem in the case when $\lambda_1\neq\lambda_2$ is that if the elliptical domain is used, (\ref{e30.20}) remains true, but $L^{\epsilon}_{P,\Theta}$ does not satisfy the spherical constraints any more, and if the spherical domain is considered, $L^{\epsilon}_{P,\Theta}$ satisfies the constraints, but (\ref{e30.20}) does not hold any longer. Therefore, in what follows, we provide some upper bounds and lower bounds for the capacity of the deterministic channel.
\begin{enumerate}
  \item Bounds based on the cubic constraints: For brevity, let
  \begin{align*}
    \mathbb{F}(\mathbf a,\mathbf b)=\{F_{\mathbf X}(\mathbf x)|&F_{X_i}(x_i)=0\mbox{ for } x_i<0, \nonumber\\&F_{X_i}(x_i)=1\mbox{ for } x_i^2\geq a_i,\nonumber\\&\int_{\mathbb{R}^n}x_i^2d^nF_{\mathbf X}(\mathbf x)\leq b_i\ ,\forall i\in[1:n]\}
  \end{align*}
  be the set of all CDFs with the cubic constraints defined by the vectors $\mathbf a$ and $\mathbf b$, respectively. By strengthening or weakening the constraints of (\ref{e2}), we have
  \begin{equation}
    \sup_{F_{\mathbf X}(\mathbf x)\in \mathbb{F}_1} I(\mathbf X;\mathbf Y)\leq C(u_p,u_a)\leq \sup_{F_{\mathbf X}(\mathbf x)\in \mathbb{F}_2} I(\mathbf X;\mathbf Y)
  \end{equation}
  as long as $\mathbb{F}_1\subseteq\{F_{\mathbf X}(\mathbf x)|F_{\mathbf X}(\mathbf x)=1 \mbox{ for } \|\mathbf x\|^2\geq u_p, \int_{\mathbb{R}^n}\|\mathbf x\|^2d^nF_{\mathbf X}(\mathbf x)\leq u_a\}\subseteq\mathbb{F}_2$. One possible choice for $\mathbb{F}_2$ is obtained with the enhanced cubic constraints as follows
  \begin{equation*}
    \mathbb{F}_2=\mathbb{F}(u_p\mathbf 1,u_a\mathbf 1)
  \end{equation*}
  where $\mathbf 1$ is the n-dimensional all-one vector. Also, a trivial option for $\mathbb{F}_1$ would be
    \begin{equation*}
    \mathbb{F}_1=\mathbb{F}(\frac{u_p}{n}\mathbf 1,\frac{u_a}{n}\mathbf 1).
  \end{equation*}
  Since the noise elements are independent, we have
  \begin{align}
    \sum_{i=1}^{n}\sup_{\substack{F_{X_i}(x_i):|X_i|^2\leq\frac{u_p}{n}\\E[|X_i|^2]\leq\frac{u_a}{n}}}\!\!\!\!\!\!\!\!\! I(X_i;Y_i)&\leq C(u_p,u_a)\nonumber\\&\leq \sum_{i=1}^{n}\sup_{\substack{F_{X_i}(x_i):|X_i|^2\leq u_p\\E[|X_i|^2]\leq u_a}}\!\!\!\!\!\!\!\!\!\!\! I(X_i;Y_i)\nonumber
  \end{align}
  which leads to
    \begin{align*}
    \sum_{i=1}^{n}C_S(\frac{\lambda_i^2u_p}{n},\frac{\lambda_i^2u_a}{n})&\leq C(u_p,u_a)\\&\leq \sum_{i=1}^{n}C_S(\lambda_i^2u_p,\lambda_i^2u_a),
  \end{align*}
  in which $C_S(.,.)$ is the capacity of a scalar AWGN channel under peak and average power constraints defined in \cite{Smith}.
  The resources could alternatively be allocated according to the noise covariance matrix $\Sigma$ such that the resource of each component is inversely proportional to its noise variance. Therefore, another possible set for obtaining a lower bound is
   \begin{equation*}
    \mathbb{F}_1=\mathbb{F}(u_p\mathbf v,u_a\mathbf v)
  \end{equation*}
  in which $v_i=\frac{\lambda_i^2}{\sum_{j=1}^{n}\lambda_j^2}$. We name this last set of constraints as modified cubic constraints.
  \item Bounds based on the elliptical constraints: Another possible set of lower and upper bounds is obtained by strengthening or weakening the constraints in (\ref{e30.11}). By noting that
  \begin{align}
    \min\{\lambda_1^{-2},\lambda_2^{-2},\ldots,\lambda_n^{-2}\}&\leq\mathbf a^T(\mathbf \theta)\mathbf\Sigma\mathbf a(\mathbf \theta)\nonumber\\&\leq\max\{\lambda_1^{-2},\lambda_2^{-2},\ldots,\lambda_n^{-2}\}\label{e.01},
  \end{align}
  we get the two following sets of constraints for the lower and the upper bounds of (\ref{e30.11}), respectively.
  \begin{align*}
    \mathbb{F}_1=\{F_{P,\mathbf \Theta}(\rho,\mathbf \theta)|&P^2\leq\min\{\lambda_1^2,\ldots,\lambda_n^2\}u_p\ ,\\&E[P^2]\leq\min\{\lambda_1^2,\ldots,\lambda_n^2\}u_a\},\\
     \mathbb{F}_2=\{F_{P,\mathbf \Theta}(\rho,\mathbf \theta)|&P^2\leq\max\{\lambda_1^2,\ldots,\lambda_n^2\}u_p\ ,\\ &E[P^2]\leq\max\{\lambda_1^2,\ldots,\lambda_n^2\}u_a\}.
  \end{align*}
  Following the same approach as in the proof of the theorem, it can be verified that with these sets of constraints, the lower and the upper bounds results from the input distributions that have finite number of concentric hyper-ellipsoids as their support.
  \item Bounds based on whitening the noise: Another trivial set of upper and lower bounds is obtained by whitening the noise and therefore, making it spherically symmetric. It is obvious that
  \begin{align}
    \sup_{\Sigma=\max\{\lambda_1^{-2},\ldots,\lambda_n^{-2}\}\mathbf I}\!\!\!\!\!\!\!\!\!I(\mathbf X;\mathbf Y)&\leq C(u_p,u_a)\nonumber\\&\leq \sup_{\Sigma=\min\{\lambda_1^{-2},\ldots,\lambda_n^{-2}\}\mathbf I}\!\!\!\!\!\!\!\!\!I(\mathbf X;\mathbf Y)\label{e.03},
  \end{align}
  where the bounds are obtained by distributions that have finite number of concentric hyper-spheres as their support as in section \ref{s2.2}. It can be easily verified that the bounds in 2) and 3) are actually the same, although the former is based on weakening or strengthening the constraint and the latter is based on whitening the noise.
  \item Lower bound based on Entropy Power Inequality (EPI): The mutual information can be lower bounded as
  \begin{align}
    I(\mathbf X;\mathbf Y)&=h(\mathbf Y) - \frac{1}{2}\ln ((2\pi e)^n|\mathbf\Sigma|)\nonumber\\
    &\geq \frac{n}{2}\ln\left(e^{\frac{2}{n}h(\mathbf X)}+e^{\frac{1}{n}\ln ((2\pi e)^n|\mathbf\Sigma|)}\right)\nonumber\\&\ \ \ - \frac{1}{2}\ln ((2\pi e)^n|\mathbf\Sigma|)\label{e.1}
  \end{align}
where in (\ref{e.1}), vector EPI \cite{network_info} has been used. In order to get a lower bound for the capacity, we notice that the maximization of $h(\mathbf X)$ under the peak and average constraints could be written as
\begin{align}
    \sup_{\substack{F_{\mathbf X}(\mathbf x):\|\mathbf X\|^2\leq u_p\\E(\|\mathbf X\|^2)\leq u_a}}\!\!\!\!\!\!h(\mathbf X)&=\!\!\!\sup_{\substack{F_P(\rho):P^2\leq u_p\\E(P^2)\leq u_a}}\!\!\!\!\!-\int_{0}^{\infty}f_P(\rho)\ln\frac{f_P(\rho)}{\rho^{n-1}}d\rho \nonumber\\&\ \ \ +\sum_{i=1}^{n-2}\ln\alpha_i + \ln 2\pi.\label{qw10}
\end{align}
By the change of variable $T=\frac{P^n}{n}$, we have
\begin{equation}\label{e.3}
    \sup_{\substack{F_P(\rho):P^2\leq u_p\\E(P^2)\leq u_a}}\!\!\!\!\!-\int_{0}^{\infty}f_P(\rho)\ln\frac{f_P(\rho)}{\rho^{n-1}}d\rho=\!\!\!\!\sup_{\substack{F_T(t):T\leq \frac{u_p^{\frac{n}{2}}}{n}\\E(T^{\frac{2}{n}})\leq \frac{u_a}{n^{\frac{2}{n}}}}}\!\!\!\!\!h(T).
\end{equation}
It can be verified that optimization theory guarantees a unique solution for (\ref{e.3}) and the necessary and sufficient conditions for $f^*_T$ to be the optimal pdf is the existence of a $\lambda\geq 0$ for which the following inequality holds for any $f_T(t)$ that has its support inside the interval $[0,\frac{u_p^{\frac{n}{2}}}{n}]$
\begin{equation}\label{e.4}
    \int_{0}^{\frac{u_p^{\frac{n}{2}}}{n}}(\ln f^*_T(t)+\lambda t^{\frac{2}{n}})(f^*_T(t)-f_T(t))dt\leq0.
\end{equation}
It is obvious that when $u_a\geq\frac{nu_p}{n+2}$, $\lambda=0$ and the optimal distribution will be uniform. In the case $u_a<\frac{nu_p}{n+2}$, $\lambda\neq0$ and the optimal distribution is given by
\begin{equation*}
    f^*_T(t) = ae^{-\lambda t^{\frac{2}{n}}}\ ,\ t\in[0,\frac{u_p^{\frac{n}{2}}}{n}],
\end{equation*}
or equivalently
\begin{equation*}
    f^*_P(\rho) = a\rho^{n-1}e^{-\frac{\lambda \rho^2}{(\sqrt[n]{n})^2}}\ ,\ \rho\in[0,\sqrt{u_p}],
\end{equation*}
since it satisfies (\ref{e.4}) with equality. The two degrees of freedom $a,\lambda$ are uniquely obtained by solving the two following equations:
\begin{align}
   &\frac{\int_{0}^{\frac{u_p^{\frac{n}{2}}}{n}}t^{\frac{2}{n}}e^{-\lambda t^{\frac{2}{n}}}dt}{\int_{0}^{\frac{u_p^{\frac{n}{2}}}{n}}e^{-\lambda t^{\frac{2}{n}}}dt} = \frac{u_a}{n^{\frac{2}{n}}}\label{e.6}\\
    &a = \left(\int_{0}^{\frac{u_p^{\frac{n}{2}}}{n}}e^{-\lambda t^{\frac{2}{n}}}dt\right)^{-1}.\label{e.7}
\end{align}
It can be verified that the left-hand side of (\ref{e.6}) is a strictly decreasing function of $\lambda$ having the range $(0,\frac{nu_p}{(n+2)n^{\frac{2}{n}}}]$ and by continuity, there exists a unique $\lambda>0$ that satisfies (\ref{e.6}). Substituting this $\lambda$ in (\ref{e.7}) gives the value of $a$ which results in
\begin{equation}\label{e.8}
    h(\mathbf X)=\frac{\lambda u_a}{(\sqrt[n]{n})^2}+\ln\left (\frac{2(\sqrt{\pi})^n}{a\Gamma(\frac{n}{2})}\right).
\end{equation}
Substituting (\ref{e.8}) in (\ref{e.1}), we get the following lower bound for the capacity
\begin{align}
    C(u_p,u_a)&\geq \frac{n}{2}\ln \left(\frac{2^{\frac{2}{n}}\pi}{(a\Gamma(\frac{n}{2}))^{\frac{2}{n}}}e^{\frac{2\lambda u_a}{n(\sqrt[n]{n})^2}}+2\pi e\sqrt[n]{|\Sigma|}\right)\nonumber\\&\ \ \ - \frac{1}{2}\ln ((2\pi e)^n|\Sigma|)\label{e.9}
\end{align}
\end{enumerate}
A visual representation of some of the bounds is shown in figure \ref{yaz} for $n=2$, $\lambda_1^2=2\lambda_2^2$ and $u_a\geq u_p$. It is obvious that the figures inside the circle (which shows the peak power constraint for the 2-dimensional channel) strengthen the constraint and those outside the circle weaken it. In figure \ref{yaz}(a), the two ellipsoids are obtained from (\ref{e.01}). In other words the inner and the outer ellipsoids are given by
\begin{equation*}
    a^T(\mathbf \theta)\mathbf\Sigma\mathbf a(\mathbf \theta)=\min\{\lambda_1^{-2},\lambda_2^{-2}\}
\end{equation*}
and
\begin{equation*}
    a^T(\mathbf \theta)\mathbf\Sigma\mathbf a(\mathbf \theta)=\max\{\lambda_1^{-2},\lambda_2^{-2}\},
\end{equation*}
respectively. The inner and outer squares in figure \ref{yaz}(b) are $[-\sqrt{\frac{u_p}{2}},\sqrt{\frac{u_p}{2}}]^2$ and $[-\sqrt{u_p},\sqrt{u_p}]^2$, respectively. The modified cubic constraint in figure \ref{yaz}(c) is based on resource allocation according to the channel gains (i.e.,$\lambda_1$ and $\lambda_2$). Channels 1 and 2 have the peak power of $\frac{\lambda_1^2}{\lambda_1^2+\lambda_2^2}u_p$($=\frac{2}{3}u_p$ in this example) and $\frac{\lambda_2^2}{\lambda_1^2+\lambda_2^2}u_p$($=\frac{1}{3}u_p$ in this example), respectively.
\section{numerical results}\label{s4}
\begin{figure*}[!t]
  \centering
  \begin{minipage}[htp]{1\textwidth}
  \subfigure[Capacity]{
    \includegraphics[width=0.4\textwidth]{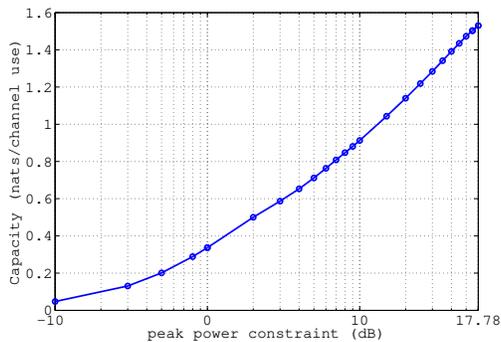}

  }
  \hfill
  \subfigure[Optimal Input Distribution]{
    \includegraphics[width=0.4\textwidth]{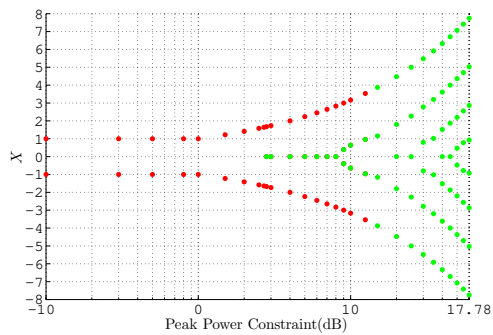}

  }

   \caption{Capacity vs. $u_p$ for $n=1$ ($u_a\geq u_p$), and the optimal input mass points.
 }\label{mass1}
\end{minipage}
\end{figure*}

\begin{figure*}[!t]
  \centering
  \begin{minipage}[htp]{1\textwidth}
  \subfigure[Capacity]{
    \includegraphics[width=0.4\textwidth]{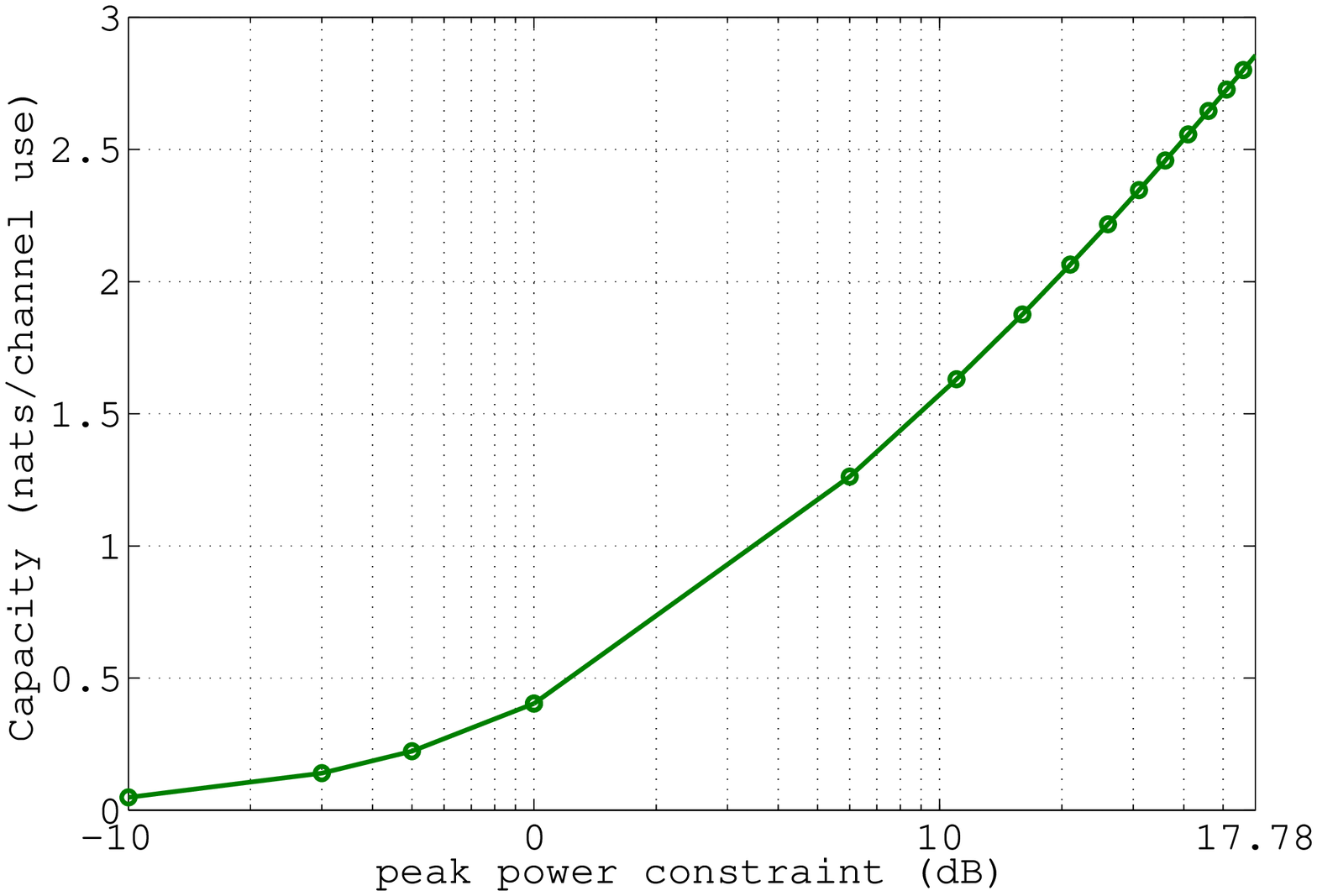}

  }
  \hfill
  \subfigure[Optimal Input Distribution]{
    \includegraphics[width=0.4\textwidth]{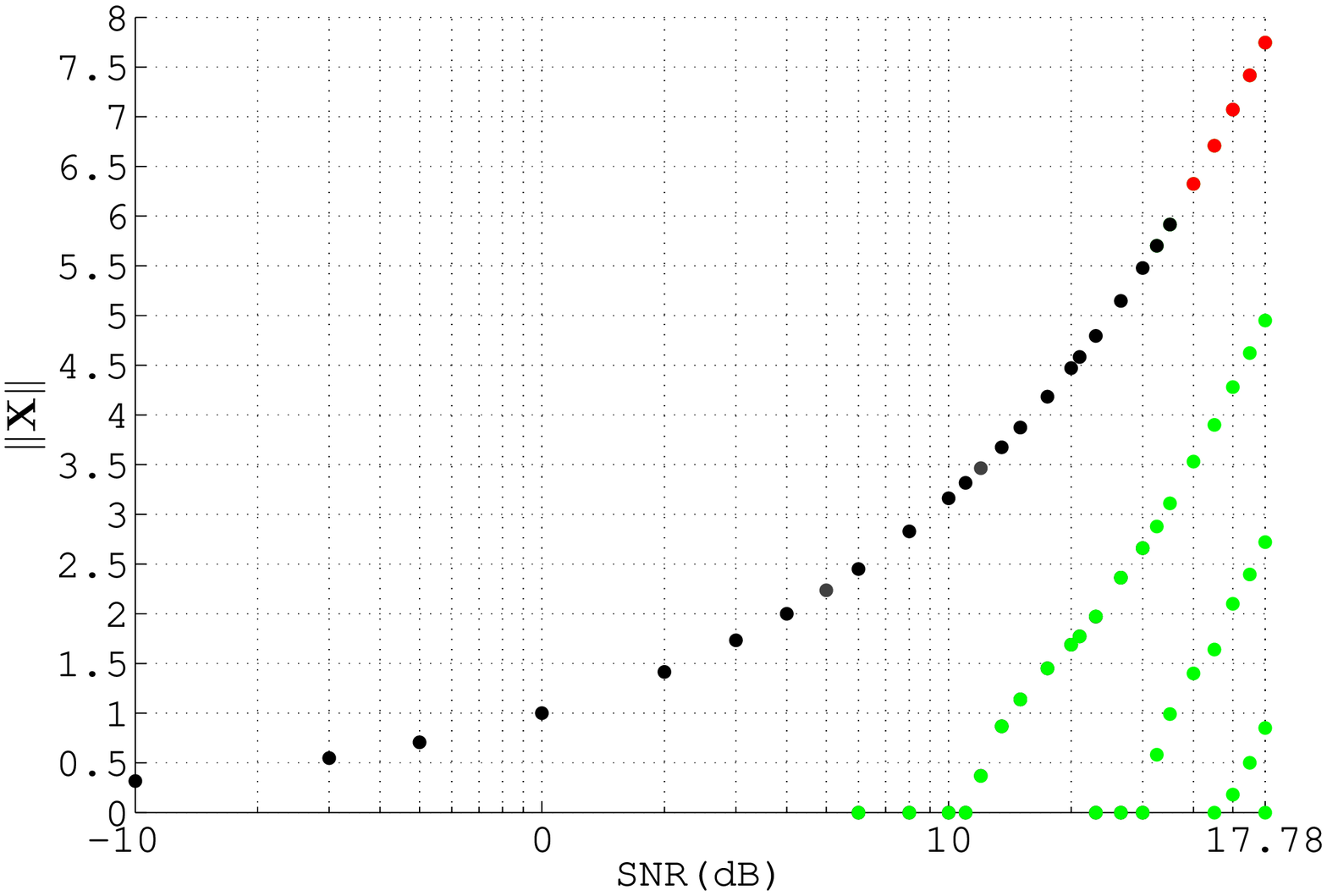}

  }

   \caption{Capacity vs. $u_p$ for $n=2$ ($u_a\geq u_p$), and the optimal input mass points.
 }\label{mass2}
\end{minipage}
\end{figure*}

\begin{figure*}[!t]
  \centering
  \begin{minipage}[htp]{1\textwidth}
  \subfigure[Capacity]{
    \includegraphics[width=0.4\textwidth]{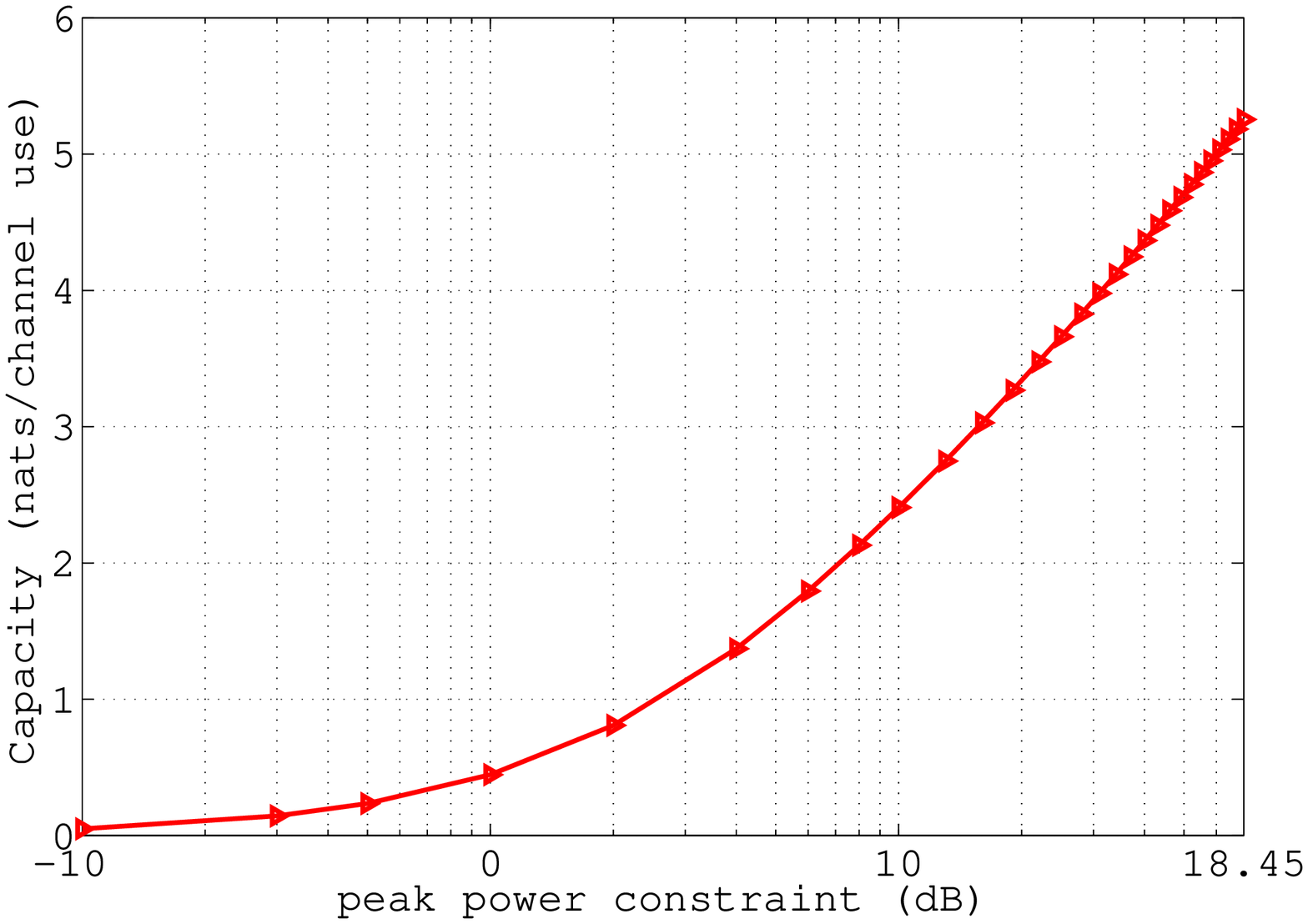}

  }
  \hfill
  \subfigure[Optimal Input Distribution]{
    \includegraphics[width=0.4\textwidth]{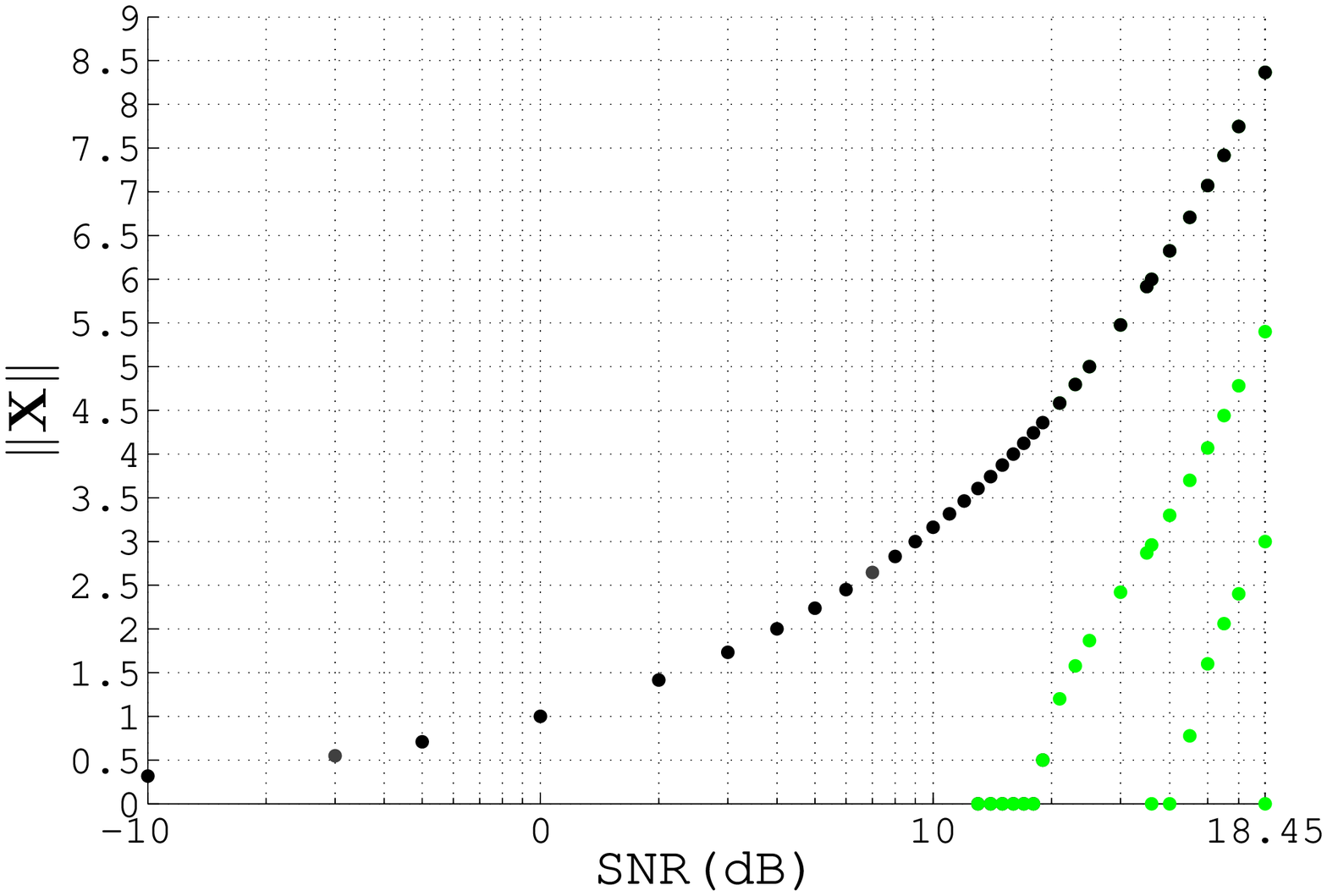}

  }

   \caption{Capacity vs. $u_p$ for $n=4$ ($u_a\geq u_p$), and the optimal input mass points.
 }\label{mass3}
\end{minipage}
\end{figure*}

\begin{figure*}[!t]
  \centering
  \begin{minipage}[htp]{1\textwidth}
  \subfigure[Capacity]{
    \includegraphics[width=0.4\textwidth]{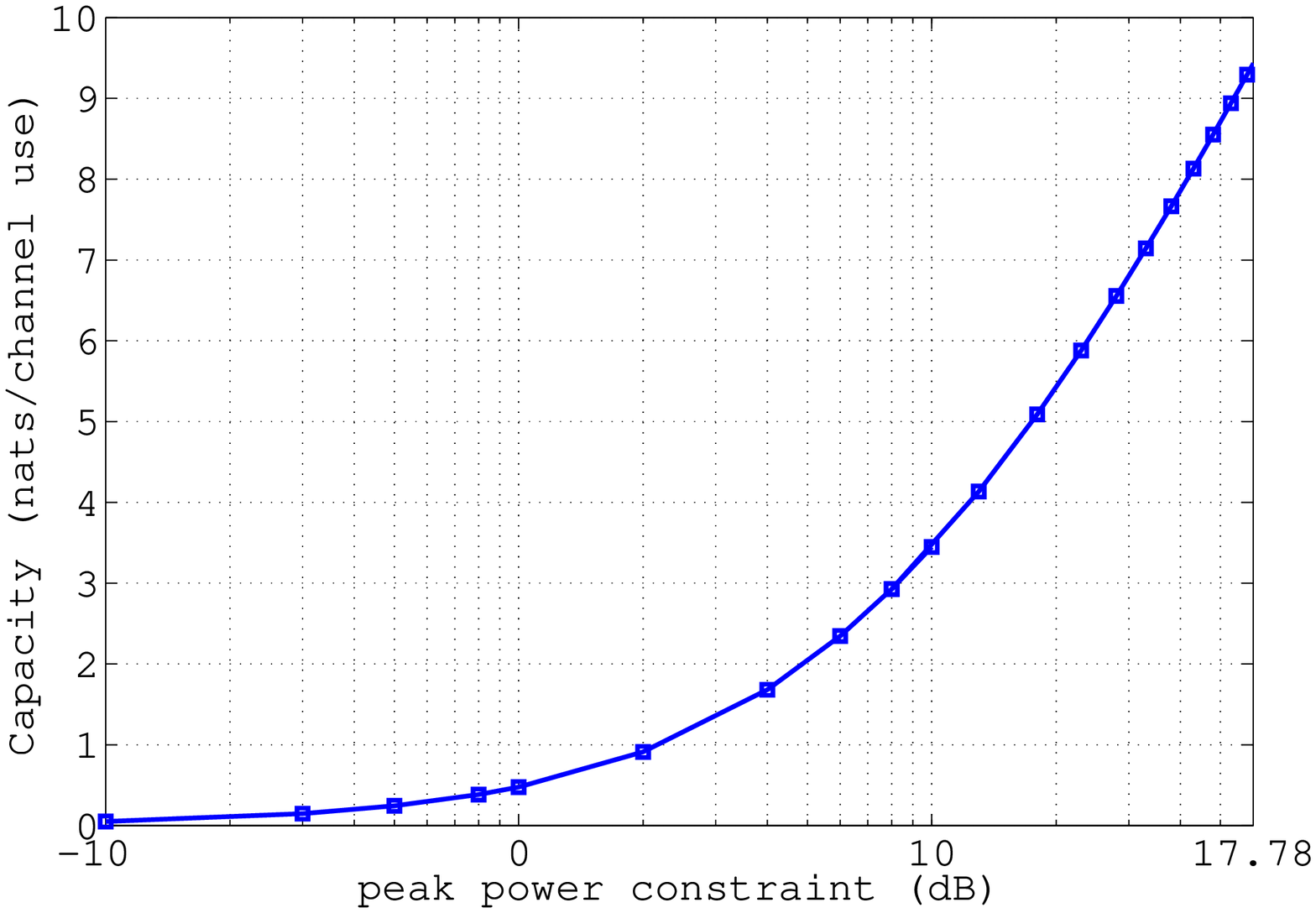}

  }
  \hfill
  \subfigure[Optimal Input Distribution]{
    \includegraphics[width=0.4\textwidth]{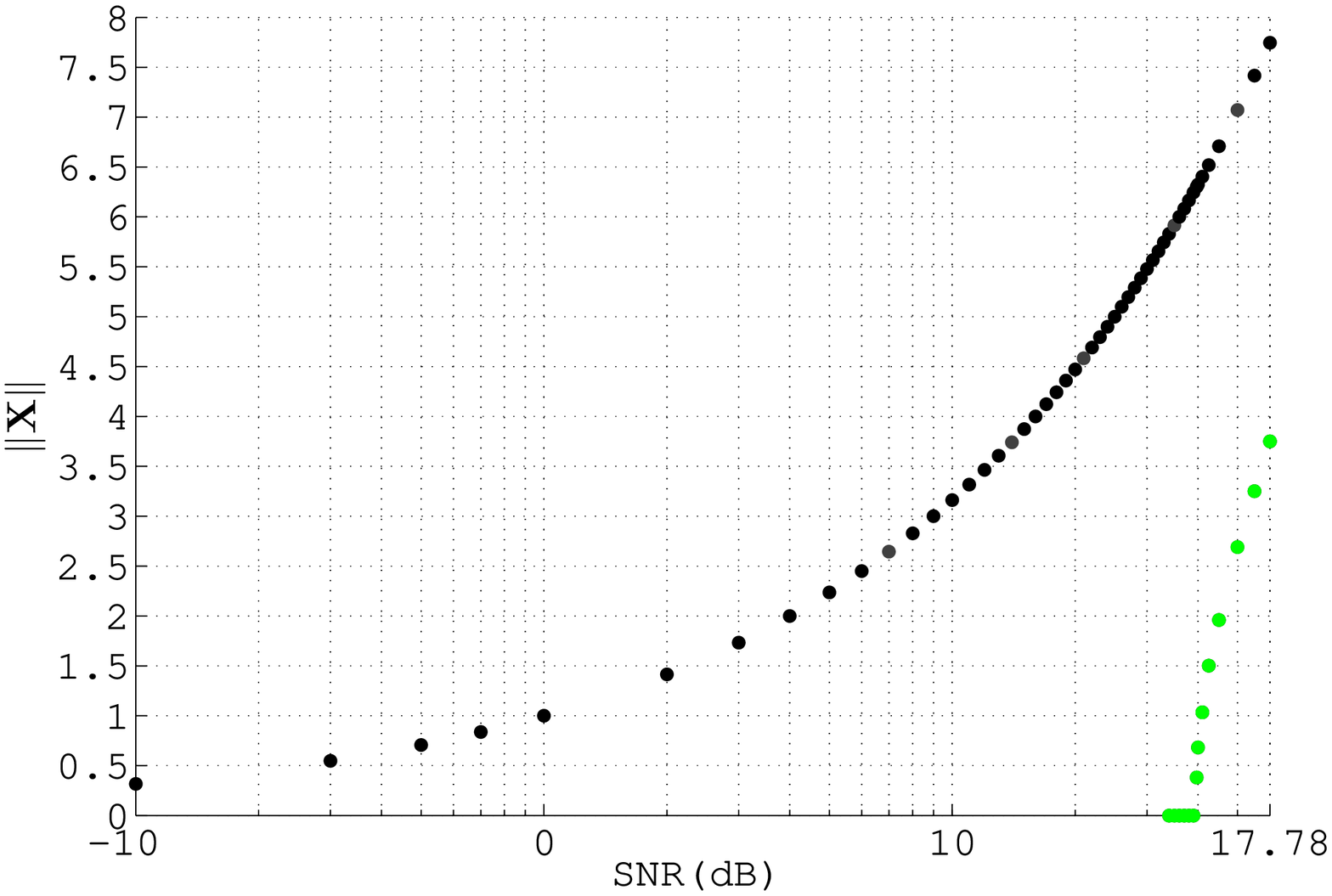}

  }

   \caption{Capacity vs. $u_p$ for $n=10$ ($u_a\geq u_p$), and the optimal input mass points.
 }\label{mass4}
\end{minipage}
\end{figure*}

\begin{figure*}[!t]
  \centering
  \begin{minipage}[htp]{1\textwidth}
  \subfigure[Capacity]{
    \includegraphics[width=0.4\textwidth]{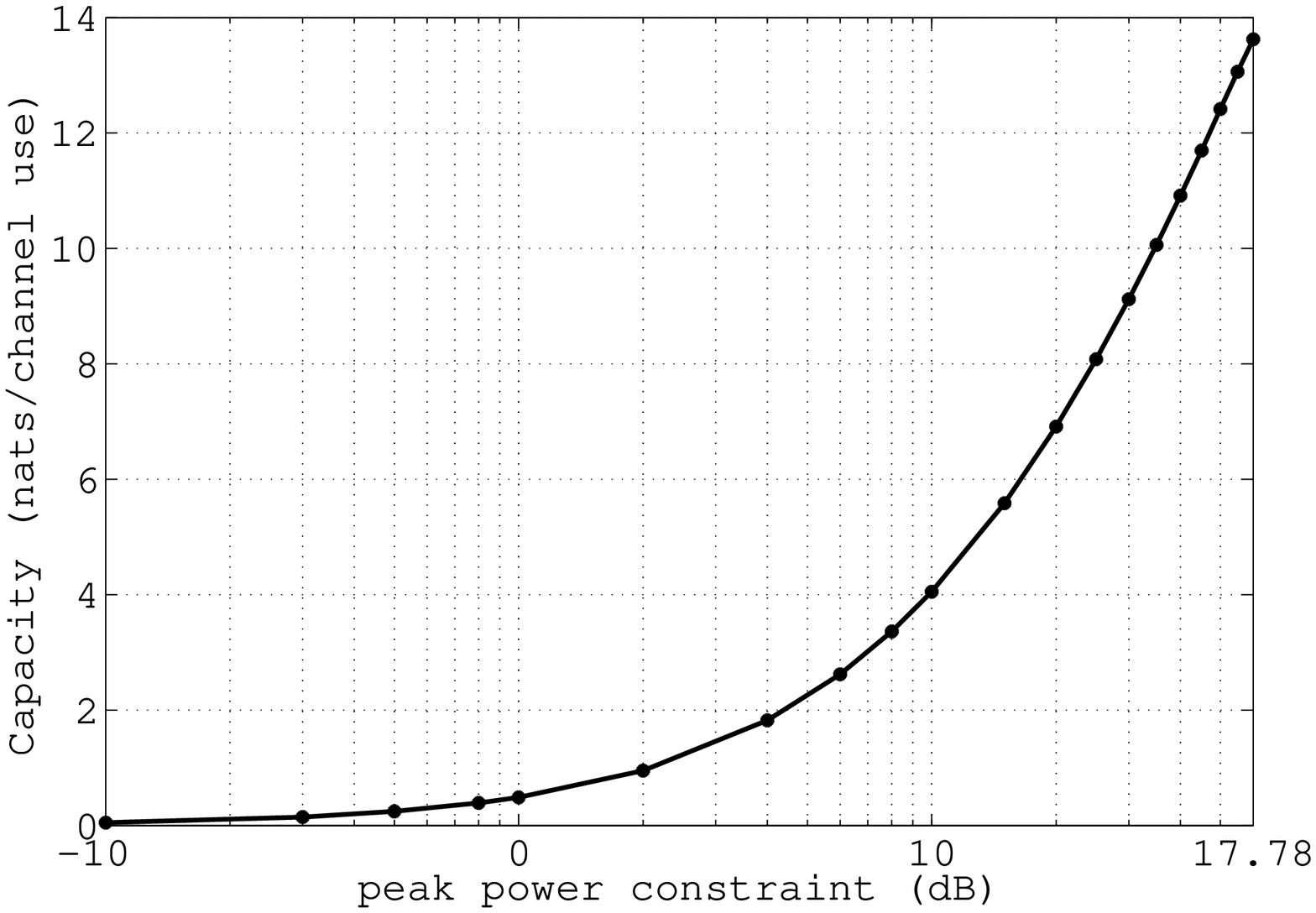}

  }
  \hfill
  \subfigure[Optimal Input Distribution]{
    \includegraphics[width=0.4\textwidth]{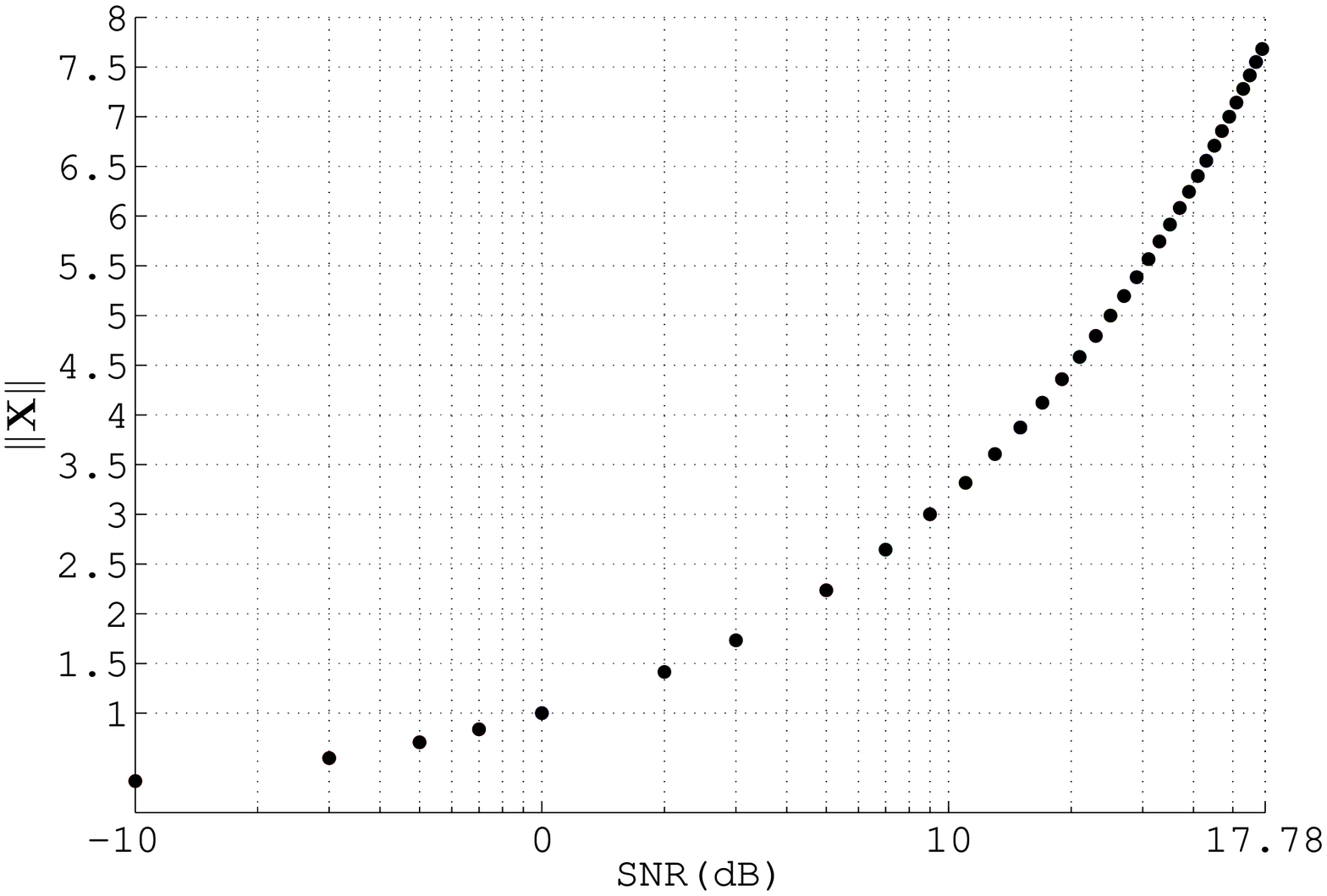}

  }

   \caption{Capacity vs. $u_p$ for $n=20$ ($u_a\geq u_p$), and the optimal input mass points.
 }\label{mass5}
\end{minipage}
\end{figure*}

As stated in the Theorem, the magnitude of the optimal input has a finite number of mass points and the phases are distributed according to (\ref{e23}). The algorithm \footnote{The codes for this section are available at http://www.ee.ic.ac.uk/bruno.clerckx/Research.html .} for finding the number, the positions and the probabilities of the optimal mass points is exactly the same as that explained in \cite{Shamai}. When the average power constraint is relaxed, figures \ref{mass1} to \ref{mass5} show the capacity of the channel in (\ref{e2}) along with the capacity-achieving input distribution for different values of $n$. In these figures, black, red and green points have their probabilities in the intervals $[0.7,1]$, $[0.3,0.7]$ and $[0,0.3]$, respectively.

\begin{figure*}[!t]
  \centering
  \begin{minipage}[htp]{1\textwidth}
  \subfigure[Capacity]{
    \includegraphics[width=0.4\textwidth]{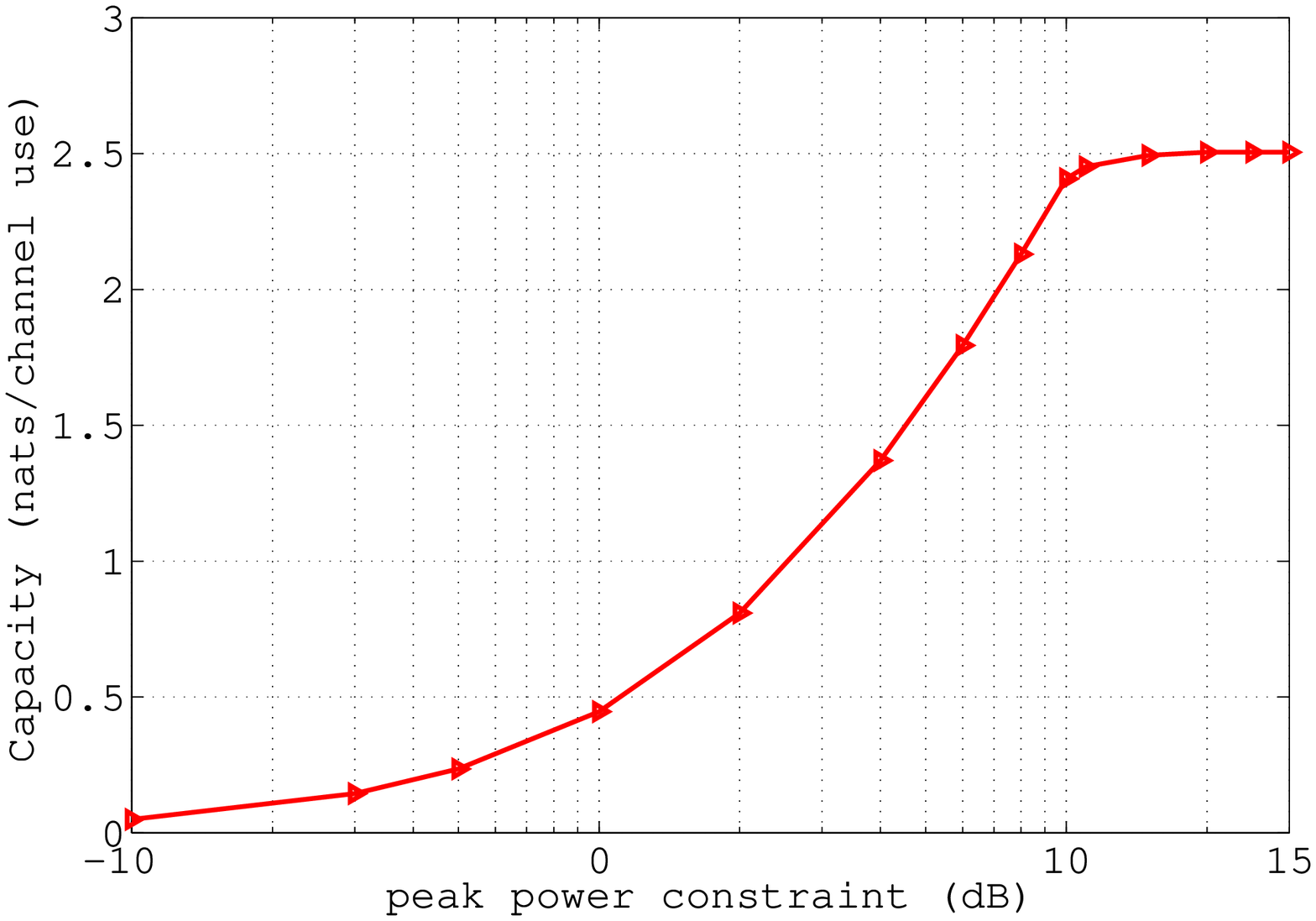}

  }
  \hfill
  \subfigure[Optimal Input Distribution]{
    \includegraphics[width=0.4\textwidth]{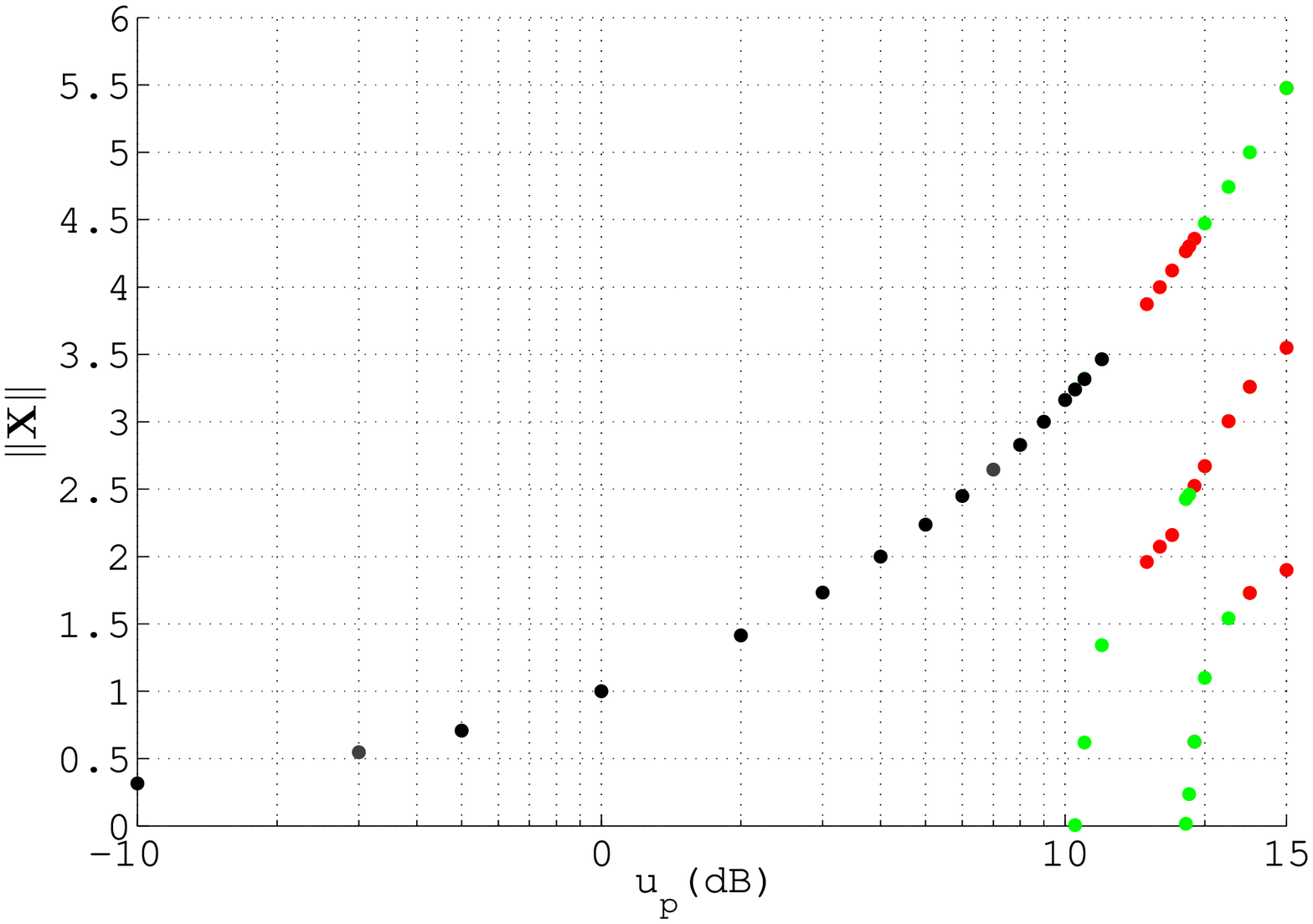}

  }

   \caption{Capacity vs. $u_p$ for $n=4$ ($u_a = 10$), and the optimal input mass points.
 }\label{mass6}
\end{minipage}
\end{figure*}

Figure \ref{mass6} shows the capacity of the four dimensional channel versus $u_p$ along with the optimal input for a fixed average power $u_a = 10$. It is obvious that the capacity saturates at its conventional value given in (\ref{e27.3}). This saturation shows the near-optimal performance of the discrete input for the conventional unbounded scenario. For example, when $n=4$ and $u_a=10$, the capacity of the channel with unbounded input (i.e., $C_G=2.5055$), which is achieved by a generalized Rayleigh distributed $P$, can also be achieved with good approximation (i.e., $I(\mathbf X;\mathbf Y)=2.5052$) by a pmf having only three mass points below $\sqrt{30}$.

\begin{figure*}[!t]
  \centering
  \begin{minipage}[htp]{1\textwidth}
  \subfigure[Capacity]{
    \includegraphics[width=0.4\textwidth]{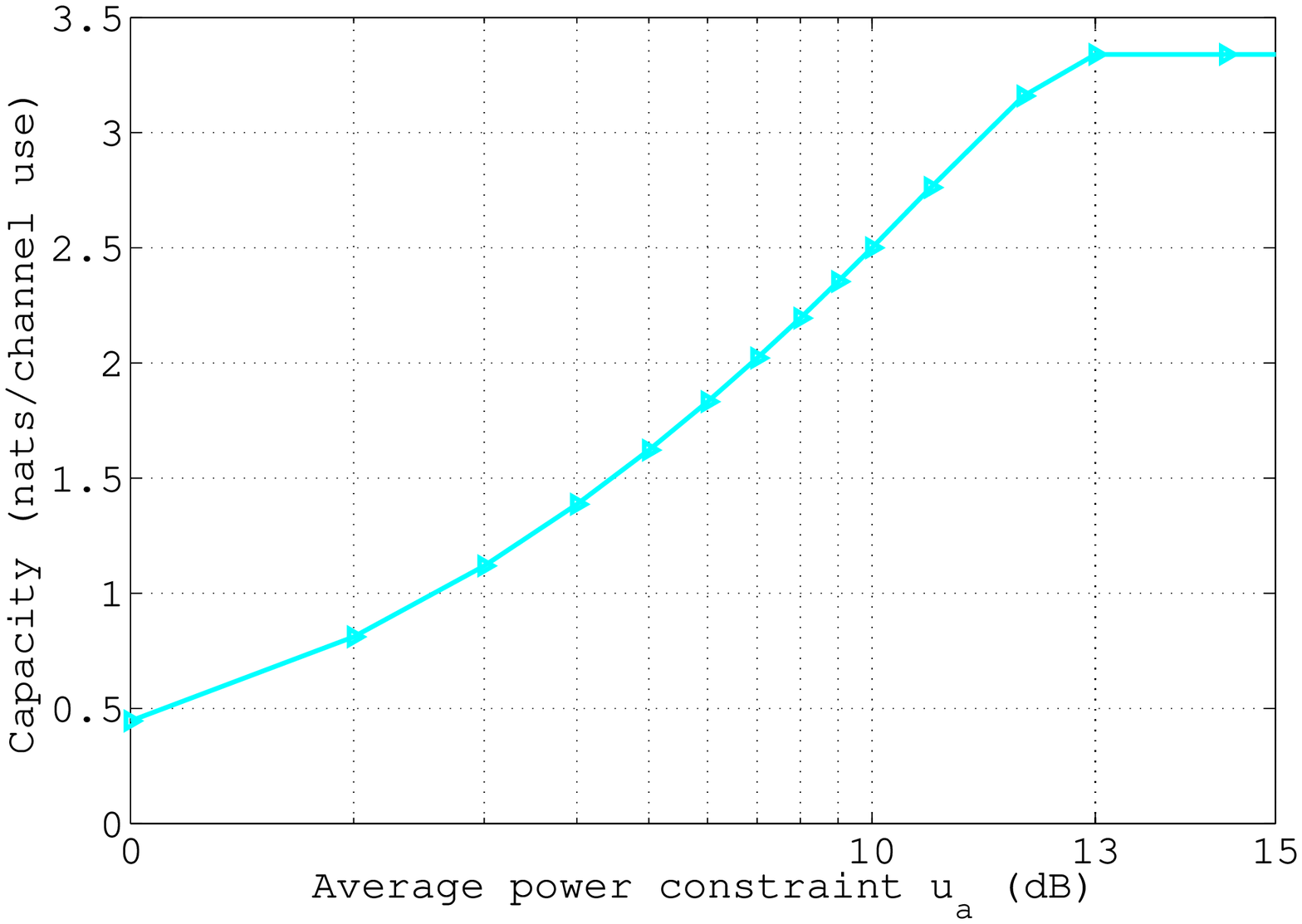}

  }
  \hfill
  \subfigure[Optimal Input Distribution]{
    \includegraphics[width=0.4\textwidth]{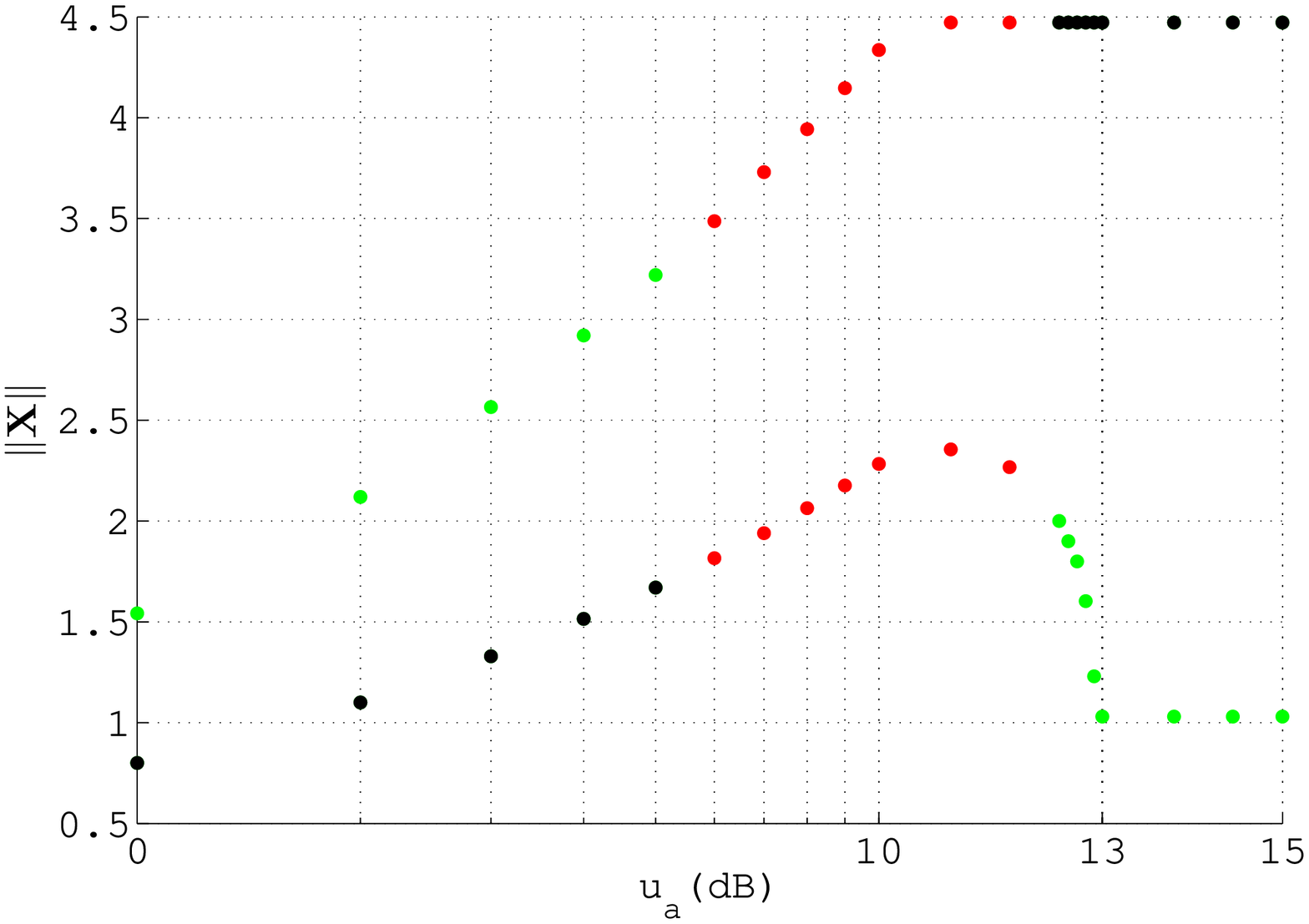}

  }

   \caption{Capacity vs. $u_a$ for $n=4$ ($u_p = 20$), and the optimal input mass points.
 }\label{mass7}
\end{minipage}
\end{figure*}

\begin{figure}[t]
  \centering
  \includegraphics[width=8cm]{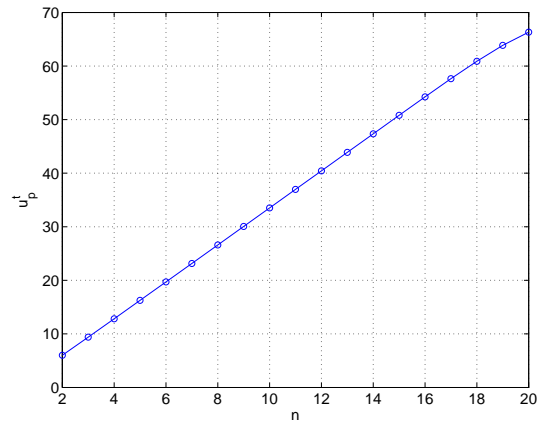}\\
  \caption{The peak power threshold for which $F_{P_1}$ remains optimal versus $n$ ($u_a\geq u_p$).}\label{fig2}
\end{figure}

Figure \ref{mass7} shows the capacity versus the average power constraint for a fixed value of the peak power ($u_p = 20$). It is obvious that for $u_a\geq u_p$, the average constraint becomes inactive and the capacity is determined only by $u_p$.
We have already shown that when the peak power is very small (i.e., $u_p\ll 1$) and $u_a\geq u_p$, the optimal input has only one mass point at $\rho=\sqrt{u_p}$. Let $F_{P_1}$ denote the cdf of this optimal input. Therefore,
\begin{align*}
    f_V(v;F_{P_1})&=K_n(v,\sqrt{u_p}),\\
    \tilde{h}_V(\rho;F_{P_1})&=-\int_{0}^{\infty}K_n(v,\rho)\ln(K_n(v,\sqrt{u_p}))dv.
\end{align*}
When $u_p\ll 1$, the above marginal entropy density is a convex and increasing function of $\rho$ and satisfies the equality of (\ref{e302}) (with $\lambda = 0$) at $\rho=\sqrt{u_p}$ and the inequality of (\ref{e301}) at all other points. As $u_p$ increases, $F_{P_1}$ remains optimal until it violates the necessary and sufficient conditions. By observing the behavior of $\tilde{h}_V(\rho,F_{P_1})$, it is concluded that as $u_p$ increases, the first point to violate the necessary and sufficient conditions will happen at $\rho=0$. Therefore, the peak power threshold $u_p^t$ for which $F_{P_1}$ remains optimal (when $u_a\geq u_p$) is obtained by solving the following equation for $u_p^t$
\begin{equation}\label{f1}
    \tilde{h}_V(0;F_{P_1})=h(V;F_{P_1}).
\end{equation}
By solving (\ref{f1}) numerically, the values of the peak power threshold are obtained for different values of $n$ as shown in figure \ref{fig2}. For example, for $n = 4$, $u_p^t\approx 12.81$ which means that when the peak power is below $12.81$, the support of the optimal input has only one hyper-sphere, and at this threshold it gets another mass point at zero as already shown in figure \ref{mass3}. For $n=20$, when $u_p\leq 66$, constant amplitude signaling is optimal which is consistent with figure \ref{mass5}. From figure \ref{fig2}, it can be observed that the ratio $\frac{u_p}{n}$ does not necessarily need to be vanishingly small to guarantee the optimality of $F_{P_1}$. Specifically, for the ratios of $\frac{u_p}{n}$ below (approximately) $3.4$, $F_{P_1}$ remains optimal.

\begin{figure}[t]
  \centering
  \includegraphics[width=8cm]{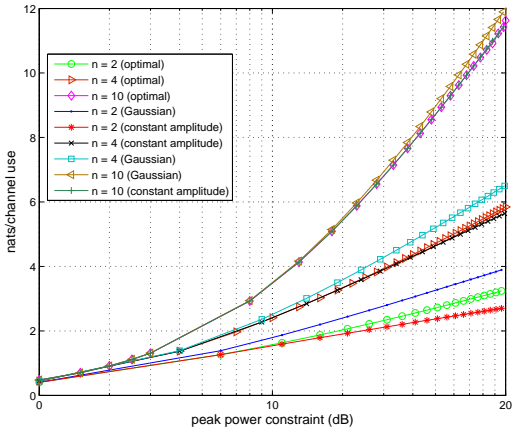}\\
  \caption{Achievable rate by the constant amplitude signaling at the peak power (i.e., $\|\mathbf X\|=\sqrt{u_p}$) when the average power constraint is relaxed.}\label{fig3.25}
\end{figure}

%
It has already been shown that when the number of antennas is above a certain threshold, constant amplitude signaling at the peak power (i.e., $\|\mathbf X\|=\sqrt{u_p}$) becomes optimal. Figure \ref{fig3.25} compares the achievable rate of the constant amplitude signaling \footnote{The rate has been obtained by numerical evaluation of
\begin{align*}
    \sup_{F_{\mathbf X}(\mathbf x):\|\mathbf X\|^2=u_p}\!\!\!\!\!I(\mathbf X; \mathbf Y)&=-\int_{0}^{\infty}e^{-\frac{(\sqrt[n]{nv})^2+u_p}{2}}\frac{I_{\frac{n}{2}-1}(\sqrt{u_p}\sqrt[n]{nv})}{(\sqrt{u_p}\sqrt[n]{nv})^{\frac{n}{2}-1}}\times\nonumber\\&\ \ \ \ln\left(e^{-\frac{(\sqrt[n]{nv})^2+u_p}{2}}\frac{I_{\frac{n}{2}-1}(\sqrt{u_p}\sqrt[n]{nv})}{(\sqrt{u_p}\sqrt[n]{nv})^{\frac{n}{2}-1}}\right)dv\\&\ \ \ -\frac{n}{2}\ln(2e)+\ln2.
\end{align*}
}at the peak power with the capacity of the channel (with the constraint $\|\mathbf X\|^2\leq u_p$) and the unbounded Gaussian input having an average power of $u_p$. As it can be observed, when the number of antennas is sufficiently large, constant amplitude signaling is not only optimal but also it has a performance close to that of the unbounded Gaussian signaling.

\begin{figure}[t]
  \centering
  \includegraphics[width=8cm]{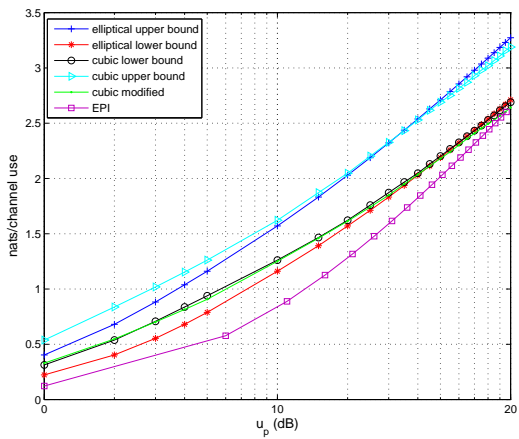}\\
  \caption{Bounds for the capacity of the deterministic MIMO channel ($\lambda_2^2 = 2\lambda_1^2 = 1$).}\label{fig4}
\end{figure}

\begin{figure}[t]
  \centering
  \includegraphics[width=8cm]{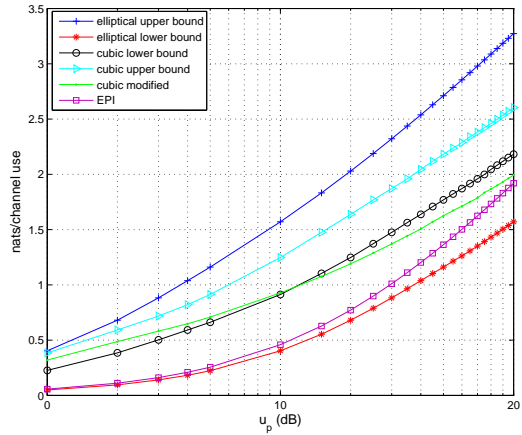}\\
  \caption{Bounds for the capacity of the deterministic MIMO channel ($\lambda_2^2 = 10\lambda_1^2 = 1$).}\label{fig5}
\end{figure}

Figures \ref{fig4} and \ref{fig5} demonstrate the bounds for the deterministic MIMO channel in (\ref{e30.01}) for two values of the condition number of the channel. It can be observed that the gap between the elliptical lower and upper bound increases with the condition number. This is intuitively justified by noting that the elliptical bounds converge to the actual capacity of the channel when the condition number approaches unity. For large values of the condition number, the lower bound obtained by modified cubic constraints performs better than the equal resource allocation at small values of the peak power.
Finally, it is important to note that although the lower bound obtained by EPI is loose in these two figures, it becomes asymptotically tight for large values of $u_p$. It can be easily verified by the fact that when the average power constraint is relaxed, we have $\lambda = 0$ and $a = \frac{n}{u_p^{\frac{n}{2}}}$ in (\ref{e.8}). When $u_p\to\infty$ the lower bound in (\ref{e.9}) gets arbitrarily close to $h(\mathbf X)$ in (\ref{e.8}) which is obviously an upper bound for the capacity. This justifies the asymptotic tightness of the bound resulted from EPI at large values of $u_p$.
\section{Conclusion}\label{s5}
We have shown that the capacity-achieving distribution of the vector Gaussian channel with identity channel matrix under the peak and average power constraints has a finite number of mass points for its amplitude and the points are uniformly distributed on the hyper-spheres determined by the amplitude mass points. It was shown that when the peak power is the only active constraint, constant amplitude signaling at the peak power is optimal when the number of dimensions is above a threshold. Finally, some upper and lower bounds were given for the general deterministic channel and their performance was evaluated numerically as a function of the condition number of the channel.

The results of the paper could be applied to the MIMO communication systems with only one single RF chain at the transmitter which is of great interest and necessitate the peak power constraint. The importance of the results becomes more pronounced in the massive MIMO settings, where it was shown that the capacity has a closed form solution and no computer program is needed to find the optimal input distribution.
\appendices
\section{derivation of (\ref{e17})}\label{a0}
The following lemma is useful in the sequel.

\textbf{Lemma 1}. Let $a$ and $b$ be two real numbers with $a>0$. Also, let $\mathbb{N}_0$ be the set of non-negative integers. Then,
\begin{align}
    \int_{-1}^{1}\!\!I_n(&a\sqrt{1-u^2})(\sqrt{1-u^2})^ne^{-bu}du\nonumber\\&=\sqrt{2\pi}a^n\frac{I_{n+\frac{1}{2}}(\sqrt{a^2+b^2})}{(\sqrt{a^2+b^2})^{n+\frac{1}{2}}},\ n=\frac{k}{2}\ ,\ \forall k\in\mathbb{N}_0.\label{e30.2}
\end{align}
\begin{proof} By using \cite[pp. 698]{Ryzhik}, (\ref{e30.2}) could be shown for $n=0$. Also, by some manipulation, (\ref{e30.2}) holds true for $n=\frac{1}{2},1,\frac{3}{2}$. For general $n$, we use induction as follows. Denote the left-hand side of (\ref{e30.2}) by $Q_n$. It is shown that if (\ref{e30.2}) is true for $n$, it will also be true for $n+\frac{1}{2}$. In other words, if
\begin{equation}\label{e30.3}
    Q_n=\sqrt{2\pi}a^n\frac{I_{n+\frac{1}{2}}(\sqrt{a^2+b^2})}{(\sqrt{a^2+b^2})^{n+\frac{1}{2}}}\ \ \ (n\geq\frac{3}{2}),
\end{equation}
then
\begin{equation}\label{e30.4}
    Q_{n+\frac{1}{2}}=\sqrt{2\pi}a^{n+\frac{1}{2}}\frac{I_{n+1}(\sqrt{a^2+b^2})}{(\sqrt{a^2+b^2})^{n+1}}.
\end{equation}
By using the recursive identity for the bessel function (i.e., $I_\alpha(z)=I_{\alpha-2}(z)-\frac{2(\alpha-1)}{z}I_{\alpha-1}(z)$), we have
\begin{align}
   Q_{n+\frac{1}{2}}&=\int_{-1}^{1}I_{n-\frac{3}{2}}(a\sqrt{1-u^2})(\sqrt{1-u^2})^{n+\frac{1}{2}}e^{-bu}du\nonumber\\&\ \ \ -\frac{2(n-\frac{1}{2})}{a}\nonumber
\end{align}
\begin{align}   
   &\ \ \ \times\int_{-1}^{1}I_{n-\frac{1}{2}}(a\sqrt{1-u^2})(\sqrt{1-u^2})^{n-\frac{1}{2}}e^{-bu}du\nonumber \\
   &=\int_{-1}^{1}I_{n-\frac{3}{2}}(a\sqrt{1-u^2})(\sqrt{1-u^2})^{n+\frac{1}{2}}e^{-bu}du\nonumber\\&\ \ \ -2(n-\frac{1}{2})\sqrt{2\pi}a^{n-\frac{3}{2}}\frac{I_{n}(\sqrt{a^2+b^2})}{(\sqrt{a^2+b^2})^{n}} \label{e30.5},
\end{align}
where in (\ref{e30.5}), we have used (\ref{e30.3}). From (\ref{e30.3}), we have
\begin{equation}\label{e30.6}
    Q_{n-\frac{1}{2}}=\sqrt{2\pi}a^{n-\frac{1}{2}}\frac{I_{n}(\sqrt{a^2+b^2})}{(\sqrt{a^2+b^2})^{n}}.
\end{equation}
By taking the derivative of (\ref{e30.6}) with respect to $a$ and using the identity $I'_\alpha(z)=\frac{1}{2}(I_{\alpha-1}(z)+I_{\alpha+1}(z))$ for $\alpha\neq0$, we have
\begin{align}
    \int_{-1}^{1}I_{n-\frac{3}{2}}(a\sqrt{1-u^2})&(\sqrt{1-u^2})^{n+\frac{1}{2}}e^{-bu}du+Q_{n+\frac{1}{2}}\nonumber\\&=2\sqrt{2\pi}\frac{\partial}{\partial a}\left\{a^{n-\frac{1}{2}}\frac{I_{n}(\sqrt{a^2+b^2})}{(\sqrt{a^2+b^2})^{n}}\right\}.\label{e30.7}
\end{align}
Solving for $Q_{n+\frac{1}{2}}$ in (\ref{e30.5}) and (\ref{e30.7}) results in
\begin{align}
    Q_{n+\frac{1}{2}}&=\sqrt{2\pi}a^{n-\frac{1}{2}}\frac{\partial}{\partial a}\left\{\frac{I_{n}(\sqrt{a^2+b^2})}{(\sqrt{a^2+b^2})^{n}}\right\}\nonumber\\
    &=\sqrt{2\pi}a^{n+\frac{1}{2}}\frac{I_{n+1}(\sqrt{a^2+b^2})}{(\sqrt{a^2+b^2})^{n+1}},\label{e30.8}
\end{align}
where in (\ref{e30.8}), we have used the identity $\frac{d}{dx}\{\frac{I_n(x)}{x^n}\}=\frac{I_{n+1}(x)}{x^n}$. This completes the proof of lemma. \qedhere
\end{proof}
(\ref{e17}) is equivalent to
\begin{align}
    \underbrace{\int_{0}^{\pi}\!\!\!\!\!\!\ldots\int_{0}^{\pi}}_{n-2\mbox{ times}}\!\!\!\int_{0}^{2\pi}\!\!\!\frac{1}{{(\sqrt{2\pi})}^n}e^{x\mathbf a^T(\mathbf \theta)\mathbf a(\mathbf \psi)}\prod_{i=1}^{n-2}\sin^{n-i-1}\psi_id\psi_{n-1}\ldots d\psi_1\nonumber\\=\left\{\begin{array}{cc} \frac{I_{\frac{n}{2}-1}(x)}{(x)^{\frac{n}{2}-1}} & x\neq0\\ \frac{1}{\Gamma(\frac{n}{2})2^{\frac{n}{2}-1}} & x=0 \end{array}\right.\forall n\geq2.\label{e30.1}
\end{align}
If $x=0$, it is obvious that the left-hand side of (\ref{e30.1}) is the hyper-surface area of an n-sphere with unit radius ($=\frac{2\pi^{\frac{n}{2}}}{\Gamma(\frac{n}{2})}$) divided by $(\sqrt{2\pi})^n$ which results in the value shown on the right-hand side. Therefore, we consider $x\neq0$. It is obvious that (\ref{e30.1}) is valid for $n=2$. Denote the left-hand side of (\ref{e30.1}) by $W_n$ and assume it is valid for $n\geq 2$. It can be verified that
\begin{align}
    W_{n+1}&=\int_{0}^{\pi}\frac{I_{\frac{n}{2}-1}(x\sin\theta\sin\psi)}{\sqrt{2\pi}(x\sin\theta\sin\psi)^{\frac{n}{2}-1}}\sin^{n-1}\psi e^{x\cos\theta\cos\psi}d\psi\nonumber\\
    &=\int_{-1}^{1}\frac{I_{\frac{n}{2}-1}(x\sin\theta\sqrt{1-u^2})}{\sqrt{2\pi}(x\sin\theta)^{\frac{n}{2}-1}}(\sqrt{1-u^2})^{\frac{n}{2}-1} e^{-x\cos\theta u}du\label{e30.9}\\
    &=\frac{I_{\frac{n-1}{2}}(x)}{(x)^{\frac{n-1}{2}}}\label{e30.91}
\end{align}
where in (\ref{e30.9}), $u=-\cos\psi$ and in (\ref{e30.91}), we have used lemma 1. This completes the proof of (\ref{e30.1}).
\section{}\label{a5}
\textbf{Proposition.} Let $X$ be a non-negative random variable and $m\in\mathbb{R}^+$. The following optimization problem
\begin{equation}\label{e64}
    \sup_{F_X(x):E[X^m]\leq A}h(X),
\end{equation}
has a unique solution. Further, the maximum is
\begin{equation}\label{e65}
   \frac{\Gamma(\frac{m+1}{m})}{\Gamma(\frac{1}{m})} -\ln\left(\frac{m\sqrt[m]{\frac{\Gamma(\frac{m+1}{m})}{\Gamma(\frac{1}{m})A}}}{\Gamma(\frac{1}{m})}\right),
\end{equation}
and is achieved by the following distribution
\begin{equation}\label{e66}
    f_{X^*}(x)=\frac{m\sqrt[m]{\frac{\Gamma(\frac{m+1}{m})}{\Gamma(\frac{1}{m})A}}}{\Gamma(\frac{1}{m})}e^{-\frac{\Gamma(\frac{m+1}{m})}{A\Gamma(\frac{1}{m})}x^m}.
\end{equation}
\begin{proof} Let $\Omega$ denote the set of all probability density functions on the non-negative real line. It can be shown that $\Omega$ is convex and compact in the Levy metric. Further, the following function
\begin{equation*}
    L(f_X(x))=h(X)-\lambda(\int_{0}^{\infty}x^mf_X(x)dx-A)
\end{equation*}
is for $\lambda\geq0$, a continuous, weakly differentiable and strictly concave function of $f_X(x)$ having the weak derivative at $f_X^0(x)$ as
\begin{equation*}
    L'_{f^0_X(x)}(f_X(x))=\int_{0}^{\infty}(\ln f_X^0(x)+\lambda x^m)(f_X^0(x)-f_X(x))dx.
\end{equation*}
Therefore, the Lagrangian optimization guarantees a unique solution for (\ref{e64}) and the necessary and sufficient condition for $f_{X^*}(x)$ to be the optimal solution is the existence of a $\lambda\geq0$ for which $L'_{f_{X^*}(x)}(f_X(x))\leq0\ \ \forall f_X(x)\in\Omega.$ It can be verified that for $\lambda=\frac{\Gamma(\frac{m+1}{m})}{A\Gamma(\frac{1}{m})}$, the distribution in (\ref{e66}) results in $L'_{f_{X^*}(x)}(f_X(x))=0$ which satisfies the necessary and sufficient conditions. Hence, the pdf in (\ref{e66}), which has the differential entropy in (\ref{e65}), is the unique solution of (\ref{e64}).
\end{proof}
\section{Proof of the theorem}\label{app}
Let $\mathbb{F}_{u_p}$ denote the space of all cumulative distribution functions satisfying the peak power constraint, i.e.
\begin{equation*}
    \mathbb{F}_{u_p}=\{F_P(\rho)|F_P(\rho)=0 \ \forall \rho<0\ ,\ F_P(\rho)=1\  \forall\rho\geq \sqrt{u_p}\}.
\end{equation*}
The metric space $(\mathbb{F}_{u_p},d_L)$ is convex and compact (\cite{Smith2}, \cite[Appendix I]{Abou}) where $d_L$ denotes the Levy metric \cite{Loeve} (note that the proof of the compactness in \cite{Abou} relies only on the average power constraint). The differential entropy $h(V;F_P):\mathbb{F}_{u_p}\to \mathbb{R}$ is continuous (\cite{Smith2}, \cite[Proposition 3]{Shamai}, \cite[Appendix I]{Abou}, \cite[Proposition 1]{Tchamkerten}) (note that the proof of continuity in \cite{Tchamkerten} is more general in the sense that it does not rely on the Schwartz properties), strictly concave and weakly differentiable (\cite{Smith2}, \cite[Proposition 4]{Shamai}, \cite[Appendix II]{Abou}, \cite[Proposition 2]{Tchamkerten}) and has the weak derivative at $F^0_P$ given by
\begin{align*}
    h'_{F^0_P}(V;F_P)&=\lim_{\zeta\to0}\frac{h(V;(1-\zeta)F^0_P+\zeta F_P)-h(V;F^0_P)}{\zeta}\nonumber\\&=\int_{0}^{\sqrt{u_p}}\!\!\tilde{h}_V(\rho;F^0_P)dF_P(\rho)\!-\!h(V;F^0_P),\forall F_P\in\mathbb{F}_{u_p}.
\end{align*}
The average power constraint is denoted by
\begin{equation*}
    G(F_P)=\int_{0}^{\sqrt{u_p}}\rho^2dF_P(\rho)-u_a\leq0.
\end{equation*}
It is obvious that $G:\mathbb{F}_{u_p}\to \mathbb{R}$ is linear and weakly differentiable having the weak derivative at $F^0_P$ given by
\begin{equation*}
    G'_{F^0_P}(F_P)=G(F_P)-G(F^0_P)\ \ ,\forall F_P\in\mathbb{F}_{u_p}.
\end{equation*}
Since $h(V;F_P)$ and $G(F_P)$ are concave maps from $\mathbb{F}_{u_p}$ to $\mathbb{R}$, Lagrangian optimization \cite{Luenberger} guarantees a unique solution for (\ref{e27}) and the necessary and sufficient condition for $F_{P^*}$ to be the optimal solution is the existence of a $\lambda(\geq0)$ such that
\begin{align}\label{e35}
    \int_{0}^{\sqrt{u_p}}(\tilde{h}_V(\rho;F_{P^*})-\lambda\rho^2)dF_P(\rho)&\leq h(V;F_{P^*})-\lambda u_a,\nonumber\\&\ \ \ \ \forall F_P\in\mathbb{F}_{u_p}.
\end{align}
It can be shown that (\ref{e35}) is equivalent to (\ref{e301}) and (\ref{e302}) (\cite[Corollary 1]{Smith}). In order to show the finiteness of the cardinality of $\epsilon_{P^*}$, we extend the marginal entropy density in (\ref{e28}) to the complex domain i.e.,
\begin{equation}\label{e36}
    \tilde{h}_V(z;F_P)=-\int_{0}^{\infty}K_n(v,z)\ln f_V(v;F_P)dv \ \ ,\ \ z\in\mathbb{C}.
\end{equation}

\textbf{Proposition 1}. The kernel $K_n(v,z)$ is an entire function in $z$ for every $v$.
\begin{proof}
This can be verified by the fact that the real and imaginary parts of $K(v,z=x+jy)$ have continuous partial derivatives and satisfy the Cauchy-Riemann equations which leads to its holomorphy over the complex plane. As a result, by Cauchy's theorem, for every rectifiable closed curve $\gamma$ in $\mathbb{C}$,
\begin{equation}\label{e36.01}
    \int_{\gamma}K_n(v,z)dz = 0.
\end{equation}\qedhere
\end{proof}
\textbf{Proposition 2}. The marginal entropy density $\tilde{h}_V(z;F_P)$ is an entire function.
\begin{proof}
First, we show the continuity of $\tilde{h}_V(z;F_P)$. Let $\{z_m\}_1^{\infty}$ be a sequence of complex numbers converging to $z_0$. Since $K_n(v,z)$ is holomorphic (see Proposition 1), it is continuous. Therefore,
\begin{equation*}
    \lim_{m\to\infty}K_n(v,z_m)\ln f_V(v;F_P)=K_n(v,z_0)\ln f_V(v;F_P).
\end{equation*}
Because the kernel is continuous and $K_n(v,+\infty)=0$, it is also bounded (i.e., $0\leq K_n(v,\rho)<\infty$ for all $\rho\in\mathbb{R}_{\geq0}.$) The continuity and boundedness of the kernel guarantees the continuity of $f_V(v;F_P)$ given in (\ref{e16}) by the application of Lebesgue's dominated convergence theorem. 
This allows us to write
\begin{align}
    0&<e^{-\frac{(\sqrt[n]{nv})^2+u_p}{2}}\frac{1}{\Gamma(\frac{n}{2})2^{\frac{n}{2}-1}}\nonumber\\&\leq\min_{\rho\in[0,\sqrt{u_p}]}K_n(v,\rho)\nonumber\\&\leq f_V(v;F_P)\nonumber\\&\leq\max_{\rho\in[0,\sqrt{u_p}]}K_n(v,\rho)\nonumber\\&\leq e^{-\frac{(\sqrt[n]{nv})^2}{2}}\frac{I_{\frac{n}{2}-1}(u_p\sqrt[n]{nv})}{(u_p\sqrt[n]{nv})^{\frac{n}{2}-1}}\nonumber\\&<\infty,\label{cimich}
\end{align}
since $\frac{I_n(x)}{x^n} (x>0)$ is a strictly increasing function. Therefore,
\begin{align}
    |\ln f_V(v;F_P)|&\leq\frac{(\sqrt[n]{nv})^2+u_p}{2}+|\ln(\frac{I_{\frac{n}{2}-1}(u_p\sqrt[n]{nv})}{(u_p\sqrt[n]{nv})^{\frac{n}{2}-1}})|\nonumber\\
    &\leq \frac{(\sqrt[n]{nv})^2+u_p}{2} + u_p\sqrt[n]{nv}+\ln(\Gamma(\frac{n}{2})2^{\frac{n}{2}-1})\label{e36.15}\\
    &\leq \frac{(\sqrt[n]{nv})^2}{2}(1+u_p)+u_p+\ln(\Gamma(\frac{n}{2})2^{\frac{n}{2}-1})\label{e36.2}
\end{align}
where in (\ref{e36.15}), we have used the inequality
\begin{equation}
    \frac{I_{\nu}(x)}{x^{\nu}}<\frac{\cosh x}{2^{\nu}\Gamma(\nu+1)}\stackrel{x>0}{<}\frac{e^x}{2^{\nu}\Gamma(\nu+1)},
\end{equation}
which was proved in \cite{Luke}. From (\ref{e36.2}), it can be verified that
\begin{align}
    &|\tilde{h}_V(z_m;F_P)|\nonumber\\&\leq\int_{0}^{\infty}|e^{-\frac{(\sqrt[n]{nv})^2+z_m^2}{2}}||\frac{I_{\frac{n}{2}-1}(z_m\sqrt[n]{nv})}{(z_m\sqrt[n]{nv})^{\frac{n}{2}-1}}||\ln f_V(v;F_P)|dv\nonumber\\
    &\leq|e^{-\frac{z_m^2}{2}}|\int_{0}^{\infty}\!\!\!e^{-\frac{(\sqrt[n]{nv})^2}{2}}\frac{I_{\frac{n}{2}-1}(|z_m|\sqrt[n]{nv})}{(|z_m|\sqrt[n]{nv})^{\frac{n}{2}-1}}|\ln f_V(v;F_P)|dv\label{bessel}\\
    &\leq|e^{\frac{|z_m|^2-z_m^2}{2}}|\left(\frac{(|z_m|^2+n)}{2}(1+u_p)+u_p\right.\nonumber\\&\ \ \ \left.+\ln(\Gamma(\frac{n}{2})2^{\frac{n}{2}-1})\right)\label{bessel2}\\
    &<\infty\nonumber
\end{align}
where in (\ref{bessel}), we have used the fact that $|I_n(z)|\leq I_n(|z|)$ and in (\ref{bessel2}) the upper bound in (\ref{e36.2}) has been used. Since the absolute value of the integrand of $\tilde{h}_V(z_n;F_P)$ is integrable, by Lebesgue's dominated convergence theorem, we have
\begin{align*}
    \lim_{m\to\infty}\tilde{h}_V(z_m;F_P)&=\lim_{m\to\infty}\int_{0}^{\infty}K_n(v,z_m)\ln f_V(v;F_P)dv\\
    &=\int_{0}^{\infty}\lim_{m\to\infty}K_n(v,z_m)\ln f_V(v;F_P)dv\\
&=\int_{0}^{\infty}K_n(v,z_0)\ln f_V(v;F_P)dv\\&=\tilde{h}_V(z_0;F_P)
\end{align*}
which proves the continuity of $\tilde{h}_V(z,F_P)$. Let $\partial T$ denote an arbitrary triangle in the complex plane. We can write,
\begin{align}
   \int_{\partial T}\tilde{h}_V(z;F_P)dz&=-\int_{\partial T}\int_{0}^{\infty}K_n(v,z)\ln f_V(v;F_P)dvdz\nonumber\\
   &=-\int_{0}^{\infty}\int_{\partial T}K_n(v,z)dz\ln f_V(v;F_P)dv\label{e36.02}\\
   &=0\label{e36.03}
\end{align}
where (\ref{e36.02}) is allowed by Fubini's theorem, because for a given rectifiable triangle $\partial T$
\begin{equation*}
    \int_{\partial T}|\tilde{h}_V(z;F_P)|dz<\infty.
\end{equation*}
(\ref{e36.03}) is due to the holomorphy of $K_n(v,z)$ (see (\ref{e36.01})). Therefore, by Morera's theorem (with weakened hypothesis), it is concluded that $\tilde{h}_V(z;F_P)$ is holomorphic on the entire complex plane.

Alternatively, the holomorphy of the marginal entropy density can be proved as follows. The following integral
\begin{equation*}
    \tilde{h}_V(z;F_P)=-\int_{0}^{\infty}K_n(v,z)\ln f_V(v;F_P)dv
\end{equation*}
is uniformly convergent for all $z\in\mathbb{K}$ (where $\mathbb{K}$ is a compact subset of $\mathbb{C}$) in the sense that for $\forall \delta>0$, there exists some real number $L_0$ such that
\begin{equation*}
   |-\int_{L_1}^{L_2}K_n(v,z)\ln f_V(v;F_P)dv|<\delta,
\end{equation*}
for $\forall L_1,L_2$ satisfying $L_0<L_1<L_2$. Therefore, by the differentiation lemma \cite{Lang}, $\tilde{h}_V(z;F_P)$ is holomorphic on the complex plane.
\end{proof}
If $\epsilon_{P^*}$ has infinite number of points, since it is a bounded subset of the real line ($\subseteq [0,\sqrt{u_p}]$), it has an accumulation point in $\mathbb{R}$ by Bolzano-Weierstrass theorem \cite{Bartle}. Hence, according to (\ref{e302}), the two holomorphic functions $\tilde{h}_V(z;F_{P^*})$ and $h(V;F_{P^*})+\lambda(z^2-u_a)$ become equal on an infinite set that has an accumulation point in $\mathbb{C}$. Therefore, by the identity theorem for holomorphic functions of one complex variable \cite{Lang}, the two functions are equal on the whole complex plane, i.e.
\begin{equation*}
    \tilde{h}_V(z;F_{P^*})=h(V;F_{P^*})+\lambda(z^2-u_a) \ \ ,\ \ \forall z\in\mathbb{C},
\end{equation*}
which results in
\begin{equation}\label{e38}
    \tilde{h}_V(\rho;F_{P^*})=h(V;F_{P^*})+\lambda(\rho^2-u_a) \ \ ,\ \ \forall \rho\in\mathbb{R}.
\end{equation}
In the following, we show that (\ref{e38}) leads to a contradiction.
\begin{enumerate}
  \item $\lambda=0$. In this case, in which the average power constraint is relaxed, (\ref{e38}) results in\\
  \begin{equation}\label{e39}
    f_V(v;F_P)=e^{-h(V;F_{P^*})},
  \end{equation}
  which is a constant and is guaranteed by the invertibility of (\ref{e28}) to be the only solution. The uniform distribution in (\ref{e39}) cannot be a legitimate pdf for $V$ on the non-negative real line. This contradiction can be observed in an alternative way. By noting that from (\ref{e39}) and (\ref{e16}), if $f_V(v;F_P)$ is to be constant (shown by $C$), then
  \begin{equation*}
    f_P(\rho)=C\rho^{n-1},\ \ \rho\geq0,
  \end{equation*}
  which is the only solution for $f_P(\rho)$ by the invertibility of (\ref{e16}). Again, it is not a legitimate pdf for $\rho$ and obviously violates the peak power constraint.
  \item $\lambda>0$. In this case (\ref{e38}) holds iff\\
  \begin{equation}\label{e41}
    f_V(v;F_P)=\frac{2(\sqrt{\lambda})^n}{\Gamma(\frac{n}{2})}e^{-\lambda(\sqrt[n]{nv})^2},
  \end{equation}
  which also holds iff
  \begin{equation}\label{e42}
   f_P(\rho)=(\sqrt{\frac{\lambda}{1-2\lambda}})^n\frac{\rho^{n-1}e^{-\frac{\lambda}{1-2\lambda}\rho^2}}{\Gamma(\frac{n}{2})},
  \end{equation}
  with $\lambda=\frac{\Gamma(\frac{n}{2}+1)}{\Gamma(\frac{n}{2})(u_a+\frac{n}{2})}.$ It is obvious that for $0<\lambda<\frac{1}{2}$, the solution in (\ref{e42}) violates the peak power constraint and for $\lambda>\frac{1}{2}$, no legitimate $f_P(\rho)$ results in (\ref{e41}). For $\lambda=\frac{1}{2}$, $f_{P}(\rho)=\delta(\rho)$ which implies a unit mass point at zero. This, of course, contradicts the first assumption of $F_{P^*}$ having infinite points of increase and also results in $C(u_p,u_a)=0.$
\end{enumerate}
Therefore, the magnitude of the optimal input has a finite number of mass points. This completes the proof of the theorem.
\section{two invertible transforms}\label{a2}
In this section, we show that the two following integral transforms are invertible (i.e., one-to-one),
\begin{align}
    q(v)&=\int_{0}^{\infty}K_n(v,\rho)t(\rho)d\rho\label{e43},\\
    w(\rho)&=\int_{0}^{\infty}K_n(v,\rho)g(v)dv\label{e44},
\end{align}
where $t$ is allowed to have at most an exponential order and $g$ a polynomial with a finite degree, so that the transforms exist. The invertibility of (\ref{e43}) and (\ref{e44}) is equivalent to the invertibility of (\ref{e16}) and (\ref{e28}), respectively. The following lemma will be helpful in the sequel.

\textbf{Lemma 2}. The kernel function $K_n(v,\rho)$ satisfies the two following equations,
\begin{align}
    \int_{0}^{\infty}K_n(v,\rho)\rho^{n-1}e^{-s\rho^2}d\rho&=\frac{e^{-\frac{s}{2s+1}(\sqrt[n]{nv})^2}}{(\sqrt{2s+1})^n}\label{e45},\\
    \int_{0}^{\infty}K_n(v,\rho)e^{-s(\sqrt[n]{nv})^2}dv&=\frac{e^{-\frac{s}{2s+1}\rho^2}}{(\sqrt{2s+1})^n}\label{e46},
\end{align}
where $s\geq0$.
\begin{proof}
From the properties of probability density functions,
\begin{equation*}
    \int_{R^n}\frac{1}{(\sqrt{2\pi\sigma^2})^n}e^{-\frac{\|\mathbf y - \mathbf x\|^2}{2\sigma^2}}d\mathbf y = 1.
\end{equation*}
By writing $\mathbf y$ and $\mathbf x$ in spherical coordinates (i.e., $\mathbf y\equiv(r,\mathbf \psi)$ and $\mathbf x\equiv(\rho,\mathbf \theta)$), and by substituting $\beta = \frac{1}{2\sigma^2}$ and $\alpha=\frac{\rho}{\sigma^2}$, we get (\ref{e48}) on top of the next page.

\begin{figure*}[!t]
\normalsize
\begin{equation}\label{e48}
    \int_{0}^{\infty}\underbrace{\int_{0}^{\pi}\ldots\int_{0}^{\pi}}_{n-2\mbox{ times}}\int_{0}^{2\pi}e^{-\beta r^2+\alpha r\mathbf a^T(\mathbf \theta)\mathbf a(\mathbf \psi)}r^{n-1}\prod_{i=1}^{n-2}\sin^{n-i-1}\psi_id\psi_{n-1}d\psi_{n-2}\ldots d\psi_1dr=(\sqrt{\frac{\pi}{\beta}})^ne^{\frac{\alpha^2}{4\beta}}.
\end{equation}
\hrulefill
\vspace*{4pt}
\end{figure*}


By using (\ref{e48}) and by change of variables, (\ref{e45}) and (\ref{e46}) are obtained. \qedhere

In order to show the invertibility of (\ref{e43}), it is sufficient to show that the following
\begin{equation}\label{e49}
    \int_{0}^{\infty}K_n(v,\rho)t(\rho)d\rho=0,
\end{equation}
results in $t(\rho)=0$. From (\ref{e49}), we have
\begin{equation*}
    \int_{0}^{\infty}\int_{0}^{\infty}K_n(v,\rho)t(\rho)d\rho e^{-s(\sqrt[n]{nv})^2}dv=0\ \ s\geq0.
\end{equation*}
By changing the order of integration, which is allowed here by Fubini's theorem, and by (\ref{e46}),
\begin{equation*}
    \int_{0}^{\infty}t(\rho)\frac{e^{-\frac{s}{2s+1}\rho^2}}{(\sqrt{2s+1})^n}d\rho =0,\ \ s\geq0,
\end{equation*}
which results in
\begin{equation}\label{e52}
    \int_{0}^{\infty}\frac{t(\sqrt{x})}{\sqrt{x}}e^{-\mu x}dx =0,\ \ \mu\in[0,\frac{1}{2}).
\end{equation}
Again, by extending $\mu$ to the complex domain, it is easy to verify that the left-hand side of (\ref{e52}) is holomorphic on the complex plane. Since this holomorphic function is zero on an infinite set ($[0,\frac{1}{2})$) which has an accumulation point in $\mathbb{C}$, it is zero on the whole complex plane and consequently the real line by the identity theorem. Therefore,
\begin{equation*}
    \int_{0}^{\infty}\frac{t(\sqrt{x})}{\sqrt{x}}e^{-\mu x}dx =0,\ \ \mu\in \mathbb{R},
\end{equation*}
which results in $t(\rho)=0.$ The uniqueness of this solution results from the invertibility of Laplace transform (by considering the non-negative values for $\mu$). It is obvious that the same approach can be carried out to show the invertibility of the transform (\ref{e44}). Alternatively, the following property of the kernel function
\begin{equation*}
    K_n(v,\rho)=K_n(\frac{\rho^n}{n},\sqrt[n]{nv})
\end{equation*}
could be used in (\ref{e43}) to show the invertibility of (\ref{e44}).
\end{proof}
\section{Alternative proof for remark 2}\label{altp}
From \cite{Karatsuba} and \cite{Alzer}, we have\footnote{Tighter bounds for Gamma function can be found in \cite{Alzer2}.}
\begin{equation}\label{gam1}
    \Gamma(x+1)<\sqrt{\pi}(\frac{x}{e})^x(8x^3 + 4x^2 + x+\frac{1}{30})^{\frac{1}{6}}.
\end{equation}
Let $f(n)\triangleq2e\left[\frac{(n-1)}{2}\Gamma(\frac{n-1}{2})\right]^\frac{2}{n-1}$. From (\ref{es}), we can write
\begin{align*}
    C_G&\geq\frac{n-1}{2}\log\left(1+\frac{u_p}{f(n)}\right)\nonumber\\
    &\geq\frac{n-1}{2}\log\left(1+\frac{u_p}{F(n)}\right),
\end{align*}
in which $F(n)$ is an upper bound for $f(n)$ and is obtained from (\ref{gam1}) as
\begin{align*}
    F(n)=2e&\left[\frac{(n-1)}{2}\sqrt{\pi}\left(\frac{n-3}{2e}\right)^\frac{n-3}{2}\left(8(\frac{n-3}{2})^3 \right.\right.\nonumber\\&\left.\left.+ 4(\frac{n-3}{2})^2 + \frac{n-3}{2}+\frac{1}{30}\right)^{\frac{1}{6}}\right]^\frac{2}{n-1}.
\end{align*}
The behavior of $F(n)$ as $n$ goes to infinity can be obtained as follows.
\begin{align*}
    \lim_{n\to\infty}\ln \frac{F(n)}{2e} &=\lim_{n\to\infty}\frac{n-3}{n-1}\ln(\frac{n-3}{2e})\nonumber\\
    &\ \ \ + \lim_{n\to\infty}\frac{2}{n-1}\ln\left[\frac{(n-1)}{2}\sqrt{\pi}\left(8(\frac{n-3}{2})^3 \right.\right.\nonumber\\&\left.\left.\ \ \ \ \ \ \ \ \ \ \ + 4(\frac{n-3}{2})^2 + \frac{n-3}{2}+\frac{1}{30}\right)^{\frac{1}{6}}\right]\nonumber\\
    & = +\infty.
\end{align*}
Therefore, $\frac{u_p}{F(n)}$ goes to zero with $n$, and from the expansion of $\ln(1+x)$ when $x\ll1$, we can write
\begin{align}
    \lim_{n\to\infty}\frac{n-1}{2}\ln\left(1+\frac{u_p}{F(n)}\right)&=\lim_{n\to\infty}\frac{u_p(n-1)}{2F(n)}\nonumber\\
    &\geq\lim_{n\to\infty}\frac{u_p(n-1)}{2(n+25)},\label{gam5}
\end{align}
where in (\ref{gam5}), we have used the fact that for $n\leq 10^{10}$, it can be verified that $n<F(n)<n+25$. The gap between $C_G$ and constant amplitude signaling can be written as
\begin{align*}
    \lim_{n\to\infty}\{C_G-\!\!\!\!\!\sup_{F_{\mathbf X}(\mathbf x):\|\mathbf X\|^2=u_p} \!\!\!\!\!I(\mathbf X;\mathbf Y)\}&\leq \lim_{n\to\infty}\frac{u_p}{2}\left(1-\frac{n-1}{n+25}\right)\\
    &=\frac{13u_p}{n+25},
\end{align*}
which completes the proof.
\section{Proof of remark 3}\label{app1}
We have
\begin{equation*}
    C(u_p,u_a)\leq C(\infty,u_a)=\frac{n}{2}\ln (1+\frac{u_a}{n}),
\end{equation*}
and
\begin{equation}\label{e59.535}
    \lim_{u_a\to 0}C(u_p,u_a)\leq \frac{u_a}{2}.
\end{equation}
The CDF $F^{**}_P(\rho)=(1-\frac{u_a}{u_p})u(\rho)+\frac{u_a}{u_p}u(\rho-\sqrt{u_p})$ induces the following output pdf
\begin{align*}
    f_V(v;F^{**}_P)&=(1-\frac{u_a}{u_p})K_n(v,0)+\frac{u_a}{u_p}K_n(v,\sqrt{u_p})\\
    &=(1-\frac{u_a}{u_p})\frac{e^{-\frac{(\sqrt[n]{nv})^2}{2}}}{\Gamma(\frac{n}{2})2^{\frac{n}{2}-1}}\nonumber\\&\ \ \ +\frac{u_a}{u_p}e^{-\frac{(\sqrt[n]{nv})^2+u_p}{2}}\frac{I_{\frac{n}{2}-1}(\sqrt{u_p}\sqrt[n]{nv})}{(\sqrt{u_p}\sqrt[n]{nv})^{\frac{n}{2}-1}}\\
    &=(1-\frac{u_a}{u_p})\frac{e^{-\frac{(\sqrt[n]{nv})^2}{2}}}{\Gamma(\frac{n}{2})2^{\frac{n}{2}-1}}\left[1+\frac{u_a}{u_p-u_a}\right.\nonumber\\&\left.\ \ \ \times\frac{e^{-\frac{u_p}{2}}\Gamma(\frac{n}{2})2^{\frac{n}{2}-1}I_{\frac{n}{2}-1}(\sqrt{u_p}\sqrt[n]{nv})}{(\sqrt{u_p}\sqrt[n]{nv})^{\frac{n}{2}-1}}\right].
\end{align*}
When $u_a$ is small, the entropy of $V$ is given by (\ref{e59.57}) on the next page.

\begin{figure*}[!t]
\normalsize

\begin{align}
    \lim_{u_a\to 0} h(V;F^{**}_P)&=\lim_{u_a\to 0}-\int_{0}^{\infty}f_V(v;F^{**}_P)\ln f_V(v;F^{**}_P)dv\\
    &=\lim_{u_a\to 0}\int_{0}^{\infty}\left\{(1-\frac{u_a}{u_p})\frac{e^{-\frac{(\sqrt[n]{nv})^2}{2}}}{\Gamma(\frac{n}{2})2^{\frac{n}{2}-1}}+\frac{u_a}{u_p}e^{-\frac{(\sqrt[n]{nv})^2+u_p}{2}}\frac{I_{\frac{n}{2}-1}(\sqrt{u_p}\sqrt[n]{nv})}{(\sqrt{u_p}\sqrt[n]{nv})^{\frac{n}{2}-1}}\right\}\times\nonumber\\
       &\ \ \ \left\{\frac{(\sqrt[n]{nv})^2}{2}+\ln\left(\frac{\Gamma(\frac{n}{2})2^{\frac{n}{2}-1}}{(1-\frac{u_a}{u_p})}\right)-\frac{u_a}{u_p-u_a}\frac{e^{-\frac{u_p}{2}}\Gamma(\frac{n}{2})2^{\frac{n}{2}-1}I_{\frac{n}{2}-1}(\sqrt{u_p}\sqrt[n]{nv})}{(\sqrt{u_p}\sqrt[n]{nv})^{\frac{n}{2}-1}}\right\}dv\label{e59.55}\\
    &=\lim_{u_a\to 0}\frac{n}{2}(1-\frac{u_a}{u_p})+(1-\frac{u_a}{u_p})\ln\left(\frac{\Gamma(\frac{n}{2})2^{\frac{n}{2}-1}}{(1-\frac{u_a}{u_p})}\right)-\frac{u_a}{u_p}+\frac{u_a}{u_p}(\frac{n+u_p}{2})\nonumber\\
    &\ \ \ +\frac{u_a}{u_p}\ln\left(\frac{\Gamma(\frac{n}{2})2^{\frac{n}{2}-1}}{(1-\frac{u_a}{u_p})}\right)-\frac{u_a^2}{u_p-u_a}\underbrace{\gamma(u_p)}_{\mbox{constant}}\label{e59.56}\\
    &= \frac{n}{2}+\frac{u_a}{2}+\ln\left(\Gamma(\frac{n}{2})2^{\frac{n}{2}-1}\right)\label{e59.57}
\end{align}
\hrulefill
\vspace*{4pt}
\end{figure*}
The six terms in (\ref{e59.56}) are obtained by multiplying the terms in the brackets of (\ref{e59.55}) in order. In (\ref{e59.57}), we have neglected the last higher order term in (\ref{e59.56}) and have used the approximation $\ln(1-x)\approx -x$ when $x\ll 1$. Therefore,
\begin{equation}\label{e59.58}
    \lim_{u_a\to 0} h(V;F^{**}_P)+\sum_{i=1}^{n-2}\ln\alpha_i+(1-\frac{n}{2})\ln2\pi-\frac{n}{2}=\frac{u_a}{2}.
\end{equation}
(\ref{e59.58}) and (\ref{e59.535}) show the asymptotic optimality of the distribution in (\ref{e59.51}).
\section{Proof of remark 4}\label{pr4}
Since $\sqrt{z}$ is holomorphic on the complex plane excluding the non-positive real line (i.e., the domain where the principal branch of the complex logarithm function is holomorphic), $g(\sqrt{x})$ has the following power series expansion about $\epsilon>0$
\begin{equation}\label{e59.06}
    g(\sqrt{x})=\sum_{m=0}^{\infty}g_m(x-\epsilon)^m=\sum_{m=0}^{\infty}\tilde{g}_mx^m,
\end{equation}
where its interval of convergence is $(0,\infty).$ Assuming infinite number of mass points, with the constraint in (\ref{e59.01}), (\ref{e38}) changes to
\begin{equation}\label{e59.02}
    \tilde{h}_V(\rho;F_{P^*})=h(V;F_{P^*})+\lambda(g(\rho)-u_a) \ \ ,\ \ \forall \rho\in\mathbb{R}
\end{equation}
or equivalently
\begin{align}
    -\int_{0}^{\infty}K_n(v,\rho)\ln f_V(v;F_P^*)dv& = \lambda g(\rho)+h(V;F_P^*)-\lambda u_a,\nonumber\\&\ \ \ \forall \rho\in\mathbb{R}.\label{e59.03}
\end{align}
Multiplying both sides of (\ref{e59.03}) by $\rho^{n-1}e^{-s\rho^2}$ ($s\geq 0$) and integrating with respect to $\rho$ gives
\begin{align*}
    -\int_{0}^{\infty}\ln f_V&(v;F_P^*)\frac{e^{-\frac{s}{2s+1}(\sqrt[n]{nv})^2}}{(\sqrt{2s+1})^n}dv \nonumber\\&= \int_{0}^{\infty}[\lambda g(\rho)+h(V;F_P^*)-\lambda u_a]\rho^{n-1}e^{-s\rho^2}d\rho,
\end{align*}
where we have used the transform in (\ref{e45}). By a change of variables as $v = \frac{t^{\frac{n}{2}}}{n}$ and $x = \rho^2$, we have
\begin{align}
    -\int_{0}^{\infty}&\ln f_V(\frac{t^{\frac{n}{2}}}{n};F_P^*)t^{\frac{n}{2}-1}\frac{e^{-\frac{s}{2s+1}t}}{(\sqrt{2s+1})^n}dt \nonumber\\&= \int_{0}^{\infty}[\lambda g(\sqrt{x})+h(V;F_P^*)-\lambda u_a]x^{\frac{n}{2}-1}e^{-sx}dx.\label{e59.05}
\end{align}
By substituting (\ref{e59.06}) in (\ref{e59.05}), we get
\begin{align*}
    -\int_{0}^{\infty}&\ln f_V(\frac{t^{\frac{n}{2}}}{n};F_P^*)t^{\frac{n}{2}-1}\frac{e^{-\frac{s}{2s+1}t}}{(\sqrt{2s+1})^n}dt \nonumber\\&= \sum_{m=1}^{\infty}\frac{\tilde{g}_m\Gamma(\frac{n}{2}+m)}{s^{\frac{n}{2}+m}}+\frac{[h(V;F_P^*)-\lambda u_a+\lambda \tilde{g}_0]\Gamma(\frac{n}{2})}{s^{\frac{n}{2}}}.
\end{align*}
Taking the inverse transform gives the unique solution as
\begin{equation*}
    \ln f_V(\frac{t^{\frac{n}{2}}}{n};F_P^*)=\sum_{m=0}^{\infty}c_mt^m,
\end{equation*}
where the coefficients are obtained from the set of equations in (\ref{e59.09}) on the next page.

\begin{figure*}[!t]
\normalsize

\begin{equation}\label{e59.09}
  \left\{\begin{array}{cc} -\sum_{m=0}^{\infty}\frac{c_m\Gamma(\frac{n}{2}+m)(2s+1)^m}{s^{\frac{n}{2}+m}}\equiv\sum_{m=1}^{\infty}\frac{\tilde{g}_m\Gamma(\frac{n}{2}+m)}{s^{\frac{n}{2}+m}}+\frac{[h(V;F_P^*)-\lambda u_a+\lambda \tilde{g}_0]\Gamma(\frac{n}{2})}{s^{\frac{n}{2}}} \\ h(V;F_P^*)=-\int_{0}^{\infty}f_V(v;F_P^*)\ln f_V(v;F_P^*)dv \end{array}\right..
\end{equation}
\hrulefill
\vspace*{4pt}
\end{figure*}

If there is no solution satisfying (\ref{e59.09}), (\ref{e59.02}) does not hold, which is the desired contradiction. However, in the case of having a solution for the coefficients in (\ref{e59.09}), we have
\begin{equation}\label{e59.10}
    f_V(v;F_P^*)=e^{\sum_{m=0}^{\infty}c_m(\sqrt[n]{nv})^{2m}}.
\end{equation}
In the case $c_m=0\ (m\geq 1)$, $f_V$ becomes a constant on the non-negative real line which cannot be a probability density function. The case $c_m=0\ (m\geq 2)$ does not result in a legitimate pdf, either (see (\ref{e41}) and its following discussion.) For the remaining case of having at least one non-zero $c_m (m\geq 3)$, (\ref{e59.10}) leads to a contradiction as follows. Let $m^*=\max_{m}\{m|c_m\neq 0\}$. If $c_{m^*}>0$, (\ref{e59.10}) is not integrable over the non-negative real line, hence, it is not a pdf. However, if $c_{m^*}<0$, no $F_P(\rho)$ can result in $f_V$, since from (\ref{cimich}),
\begin{equation}\label{e59.11}
    f_V^{-1}(v;F_P)=O(e^{\frac{(\sqrt[n]{nv})^2}{2}}),
\end{equation}
while the behavior of the inverse of (\ref{e59.10}) is different from (\ref{e59.11}) as $v$ goes to infinity. Therefore, it is concluded that (\ref{e59.10}) cannot be resulted by any $F_P(\rho)$ due to its behavior at large $v$. This implies that the discrete nature of the magnitude of the optimal input distribution does not change when the average constraint is generalized to (\ref{e59.01}).
\section*{Acknowledgement}
Borzoo Rassouli would like to thank M. Sedaghat and A. Tchamkerten for helpful discussions.
\bibliography{REFERENCE}

\begin{thebibliography}{10}
\providecommand{\url}[1]{#1}
\csname url@samestyle\endcsname
\providecommand{\newblock}{\relax}
\providecommand{\bibinfo}[2]{#2}
\providecommand{\BIBentrySTDinterwordspacing}{\spaceskip=0pt\relax}
\providecommand{\BIBentryALTinterwordstretchfactor}{4}
\providecommand{\BIBentryALTinterwordspacing}{\spaceskip=\fontdimen2\font plus
\BIBentryALTinterwordstretchfactor\fontdimen3\font minus
  \fontdimen4\font\relax}
\providecommand{\BIBforeignlanguage}[2]{{%
\expandafter\ifx\csname l@#1\endcsname\relax
\typeout{** WARNING: IEEEtran.bst: No hyphenation pattern has been}%
\typeout{** loaded for the language `#1'. Using the pattern for}%
\typeout{** the default language instead.}%
\else
\language=\csname l@#1\endcsname
\fi
#2}}
\providecommand{\BIBdecl}{\relax}
\BIBdecl

\bibitem{Smith}
J.~Smith, ``The information capacity of amplitude and variance constrained
  scalar gaussian channels,'' \emph{Inform. Contr.}, vol.~18, pp. 203--219,
  1971.

\bibitem{Shamai}
S.~Shamai and I.~Bar-David, ``The capacity of average and peak-power-limited
  quadrature gaussian channels,'' \emph{IEEE Trans. Inf. Theory}, vol.~41,
  no.~4, pp. 1060--1071, July 1995.

\bibitem{Abou}
I.~Abou-Faycal, M.~Trott, and S.~Shamai, ``The capacity of discrete-time
  memoryless rayleigh-fading channels,'' \emph{IEEE Trans. Inf. Theory},
  vol.~47, no.~4, pp. 1290--1300, May 2001.

\bibitem{Katz}
M.~Katz and S.~Shamai(Shitz), ``On the capacity-achieving distribution of the
  discrete-time noncoherent and partially coherent awgn channels,'' \emph{IEEE
  Trans. Inf. Theory}, vol.~50, no.~10, pp. 2257--2270, October 2004.

\bibitem{Gursoy}
M.~Gursoy, H.~Poor, and S.~Verd\'{u}, ``The noncoherent rician fading
  channel-parti: Structure of the capacity-achieving input,'' \emph{IEEE Trans.
  Wireless Comm.}, vol.~4, no.~5, pp. 2193--2206, September 2005.

\bibitem{Tchamkerten}
A.~Tchamkerten, ``On the discreteness of capacity-achieving distributions,''
  \emph{IEEE Trans. Inf. Theory}, vol.~50, no.~11, pp. 2773--2778, November
  2004.

\bibitem{khandani}
B.~Mamandipoor, K.~Moshkar, and A.~Khandani, ``Capacity-achieving distributions
  in {G}aussian multiple access channel with peak power constraints,''
  \emph{IEEE Trans. Inf. Theory}, vol.~60, no.~10, pp. 6080--6092, October
  2014.

\bibitem{Sedaghat}
M.~A. Sedaghat, R.~R. Mueller, and G.~Fischer, ``A novel single-rf transmitter
  for massive mimo,'' in \emph{18th International ITG Workshop on Smart
  Antennas (WSA)}.\hskip 1em plus 0.5em minus 0.4em\relax VDE, 2014, pp. 1--8.

\bibitem{Palanki}
R.~Palanki, ``On the capacity achieving distributions of some fading
  channels,'' in \emph{Proc., 40th Annu. Allerton Conf. Communication, Control,
  and Computing}, Monticello, IL, Oct. 2002, pp. 337--346.

\bibitem{Chan}
S.~Chan, S.~Hranilovic, and F.~Kschischang, ``Capacity-achieving probability
  measure for conditionally gaussian channels with bounded inputs,'' \emph{IEEE
  Trans. Inf. Theory}, vol.~51, no.~6, pp. 2073--2088, June 2005.

\bibitem{Sommerfeld}
J.~Sommerfeld, I.~Bjelakovi\'{c}, and H.~Boche, ``On the boundedness of the
  support of optimal input measures for rayleigh fading channels,'' in
  \emph{IEEE International Symposium on Information Theory (ISIT)}, July 2008,
  pp. 1208--1212.

\bibitem{Telatar}
E.~Telatar, ``Capacity of multi-antenna gaussian channels,'' \emph{European
  Trans. on Telecommunications}, vol.~10, no.~6, pp. 585 --595, 1999.

\bibitem{network_info}
A.~E. Gamal and Y.-H. Kim, \emph{Network Information Theory}.\hskip 1em plus
  0.5em minus 0.4em\relax Cambridge University Press, 2012.

\bibitem{Ryzhik}
I.~S. Gradshteyn and I.~M. Ryzhik, \emph{Table of integrals, series, and
  products}.\hskip 1em plus 0.5em minus 0.4em\relax Academic Press, 2007.

\bibitem{Smith2}
J.G.Smith, \emph{On the information capacity of peak and average power
  constrained Gaussian channels}.\hskip 1em plus 0.5em minus 0.4em\relax Ph.D.
  dissertation., Dep. Elec. Eng., Univ. of California, Berkeley, CA, Dec. 1969.

\bibitem{Loeve}
M.~Lo\'{e}ve, \emph{Probability Theory}.\hskip 1em plus 0.5em minus 0.4em\relax
  New York:Van Nostrand, 1960.

\bibitem{Luenberger}
D.~G. Luenberger, \emph{Optimization by Vector Space Methods}.\hskip 1em plus
  0.5em minus 0.4em\relax New York:Wiley, 1969.

\bibitem{Luke}
Y.~Luke, ``Inequalities for generalized hypergeometric functions,''
  \emph{Journal of Approximation Theory}, vol.~5, no.~1, pp. 41--65, January
  1972.

\bibitem{Lang}
S.~Lang, \emph{Complex Analysis}.\hskip 1em plus 0.5em minus 0.4em\relax 4th
  ed. Springer, 1999.

\bibitem{Bartle}
R.~Bartle, \emph{Elements of Real Analysis}.\hskip 1em plus 0.5em minus
  0.4em\relax New York:Wiley, 1964.

\bibitem{Karatsuba}
E.~Karatsuba, ``{O}n the asymptotic representation of the {E}uler gamma
  function by {R}amanujan,'' \emph{J. Computat. Appl. Math}, vol. 135, pp.
  225--240, 2001.

\bibitem{Alzer}
H.~Alzer, ``{O}n {R}amanujan's double-inequality for the gamma function,''
  \emph{Bull. Lond. Math. Soc.}, vol.~35, pp. 601--607, 2003.

\bibitem{Alzer2}
------, ``{S}harp upper and lower bounds for the gamma function,'' \emph{Royal
  Society of Edinburgh}, vol. 139A, p. 709–718, 2009.

\end{thebibliography}
\bibliographystyle{IEEEtran}
\begin{IEEEbiographynophoto}{Borzoo Rassouli}
received his M.Sc. and Ph.D. degrees in communication systems engineering from University of Tehran, Iran, in 2012 and Imperial College, London, UK in 2016, respectively. He is currently a research associate in the Intelligent Systems and Networks Group at Imperial College. His research interests are in the general area of information theory, wireless communications, detection and estimation theory.
\end{IEEEbiographynophoto}

\begin{IEEEbiographynophoto}{Bruno Clerckx}
received the M.S. and Ph.D. degrees
in applied science from Universite catholique de
Louvain, Belgium. He is now a Senior Lecturer (Associate
Professor) at Imperial College London. He held visiting
research positions at Stanford University and
EURECOM and was with Samsung Electronics from
2006 to 2011. He actively contributed to 3GPP LTE/
LTE-A and IEEE802.16m.
He is the author or coauthor of two books on
MIMO wireless communications and networks and
numerous research papers, standard contributions
and patents. He received the Best Student Paper Award at the IEEE Symposium
on Communications and Vehicular Technology in 2002 and several awards from
Samsung in recognition of special achievements. Dr. Clerckx has served as an
Editor for IEEE TRANSACTIONS ON COMMUNICATIONS and is currently an Editor for IEEE TRANSACTIONS ON WIRELESS COMMUNICATIONS.
\end{IEEEbiographynophoto}
\end{document}